\documentclass[fleqn,twoside,onecolumn,nofootinbib,showkeys,11pt]{revtex4}
\usepackage{graphicx}

\usepackage{bm}% bold math

\begin{document}
%

%--------------------------   Title   -------------------------------%
\begin{center}
{\bf 
A possible  evidence  of  the  hadron-quark-gluon  mixed phase formation in nuclear collisions
}

{\textbf V.A.$\,$Kizka$^{a,}$\footnote{\normalfont Corresponding author E-mail address:
Valeriy.Kizka@cern.ch}, V.S.$\,$Trubnikov$^{b}$, K.A. Bugaev$^{c}$ and D.R. Oliinychenko$^{c, d}$}

\emph{\small $^a$
V. N. Karazin Kharkiv National University, 61022  Kharkov, Ukraine}

\emph{\small $^b$National Science Center ``Kharkov Institute of
Physics and Technology", 61108  Kharkov, Ukraine}

\emph{\small $^c$Bogolyubov Institute for Theoretical Physics, National Academy of Sciences of Ukraine, 03680 Kiev,  Ukraine}

\emph{\small $^d$FIAS, Goethe University,  Ruth-Moufang Str. 1, 60438 Frankfurt upon Main, Germany}

\end{center}
%--------------------------------------------------------------------%

\date{}

\thispagestyle{empty}
\begin{center}
\begin{minipage}{160.7mm}
{\small
%----------------------   Abstract  --------------------------------%

{\bf Abstract.} The  performed   systematic meta-analysis of the quality of data description (QDD)
of  existing event generators of nucleus-nucleus collisions 
allows us to extract a very  important physical information. 
Our meta-analysis is dealing with the results of 10 event generators which describe   data measured in the range of center of mass collision energies from 3.1 GeV to 17.3 GeV. 
It considers the mean deviation squared per  number  of experimental points  obtained by these event generators, i.e.
the QDD,  as the results of independent  meta-measurements. 
These generators and their QDDs 
are divided in two groups. The first group includes the generators which  account for 
the   quark-gluon plasma formation  during nuclear collisions  (QGP models), 
while the second group includes the generators which 
do not assume the   QGP  formation in such collisions (hadron gas models).  
Comparing the QDD of  more than a hundred of different data sets  of strange hadrons by two groups of models, we found two regions of the equal quality description  of  data which are located  at the 
center of mass collision energies 4.4-4.87 GeV and 10.8-12 GeV. At the collision energies below 4.4 GeV the hadron gas models describe data much better  than the QGP one and, hence,  we associate this region with hadron phase.
At  the collision energies between 5 GeV and 10.8 GeV and above 12 GeV we found that QGP models describe data essentially better
than the hadron gas ones and, hence, these regions we associate with the quark-gluon phase.  
As a result,   the collision energy  regions 4.4-4.87 GeV and 10.8-12 GeV we interpret as the energies of the  hadron-quark-gluon mixed phase formation.
Based on these findings we argue  that the most probable energy range of the QCD phase diagram  (tri)critical endpoint is 
12-14 GeV.  The practical suggestions for the collision energies of the future RHIC Beam Energy Scan Program are made.
}\\
PACS numbers: 25.75.-q, 03.65.Pm, 03.65.Ge, 61.80.Mk\\
%-------------------------------------------------------------------%
\end{minipage}
\end{center}
%--------------------  Text of the article  ------------------------%

\section{INTRODUCTION}

The correct determination of the threshold energy  of Quark Gluon Plasma formation (QGP) is one of the major  goals of  modern  heavy ion   physics.  The experiments planned  at FAIR \cite{Bleich11} and NICA \cite{Kolesnik13} are targeted  to study the properties of  exotic nuclear matter  created  in 
nucleus-nucleus  (A+A) collisions and to figure out a precise  position of the  critical endpoint of the
deconfinement  phase transition \cite{Fukushim11}. 
The
Beam Energy Scan  program performed  at RHIC revealed  somewhat different  properties
of strongly interacting  matter created  in  A+A collisions at energies below and above 
$\sqrt{s_{NN}}\,=\,11.5$ GeV \cite{Mohanty13}. 
At the same time,  the  experiments on central heavy ion collisions  done at  AGS and SPS  
found such remarkable irregularities as the
Strangeness  Horn in $K^{+}/\pi^{+}$ ratio at $\sqrt{s_{NN}}\,=\,7.6$ GeV  \cite{Alt08} 
and the peaks of  $\bar{\Lambda}/\bar{p}$ and $\Lambda/\pi^-$ ratios at  energies about 
$\sqrt{s_{NN}}\,=\,5-6 $ GeV  \cite{Wang12, Alt008}.
Although there exist  some claims \cite{Horn, Step, Gazd_rev:10} that the Strangeness  Horn signals about the onset of deconfinement, the convincing explanation of its relation to the deconfinement is, in fact, absent. 
Despite the fact that a very successful  description  of  the $K^{+}/\pi^{+}$ and $\Lambda/\pi^-$ peaks  is achieved recently  in  
\cite{Bugaev13:Horn,Bugaev13:SFO,Sagun14:Lambda}, such a success  does not provide  an  apparent relation  between  the strangeness enhancement  at these collision energies and the  deconfinement transition.

Moreover, nowadays not only the situation with the ``signals" of QGP formation  is unclear, but also there is no consensus on its formation threshold energy.  For example, the authors of Strangeness Horn ``signal''  \cite{Horn}, claim that the onset of deconfinement starts at $\sqrt{s_{NN}}\,=\,7.6$ GeV  \cite{Gazd_rev:10}. 
On the other hand, the authors of  \cite{RafLet} state that the  onset of deconfinement occurs  at $\sqrt{s_{NN}}\,=\,6.3$ GeV, while very recently there appeared the  works \cite{Bugaev:2014a, Bugaev:2014b} which are arguing that the hadron-QGP mixed  phase is formed at $\sqrt{s_{NN}}\,=\,4.2 - 4.9$ GeV.
Regarding these threshold energy values,  we would like to stress that the  works \cite{Horn, Step, Gazd_rev:10,RafLet}  do not provide the  underling statistical or hydrodynamic or microscopic  models which are able to connect their ``signals"  with
the very  fact of  the hadron-QGP mixed  phase  formation. In Refs.  \cite{Bugaev:2014a, Bugaev:2014b}  the underlying hydrodynamic model of phase transition which explains  the observed irregularities is developed, 
but  similarly to the   works  mentioned above, i.e.  \cite{Gazd_rev:10, RafLet},   
the  model of   Refs.  \cite{Bugaev:2014a, Bugaev:2014b}  is  also  relying only on a certain  set of   experimental data from many sets of existing  data. 

At the same time various sets of   experimental data (the transverse mass or transverse momentum distributions, the longitudinal 
rapidity distributions of several hadron species,  the total or/and midrapidity yields of various hadrons e.t.c.)  are, in principle,  available and the results of a few event generators of A+A collisions are published. 
These generators can be divided into two major groups: the first group 
includes the  QGP state in their description (QGP models), while the second group does not include the QGP formation, i.e. Hadron Gas (HG)  models. 
If the existing generators of  A+A collisions contain just a grain of truth, then,  we believe, it is worth to systematically  compare their ability to reproduce experimental data and to  find out this grain from such a comparison. 
Hence, in this work we suggest to employ  both the available sets of  data and the theoretical models describing them
in order to elucidate the correct value of the threshold energy of  the   hadron-QGP mixed  phase.

{Our main idea is based on the assumption  that  the HG models of  A+A collisions should provide 
worse description of the  data above  the  QGP threshold energy,  whereas below this threshold they should be able to better (or at least not worse) reproduce data compared to the QGP models. 
Furthermore, we assume  that  both kinds of models should provide an equal  and rather  good QDD at the 
energy  of mixed phase production. Hence, the mixed phase production threshold should be slightly below the  energy 
at which the equal QDD  is changed to the essential worsening of QDD  by HG  models.}

The second primary aim of this work is to make a comparative study of two classes of the event generators of  A+A collisions.
This is necessary to fix the present days status quo,  to make a realistic plan of further experiments on A+A collisions   and to formulate the most important tasks for theoretical studies. 

The work is organized as follows. In the next section we present the basic elements of our meta-analysis.
Section III is devoted to a detailed description of two ways of averaging used in this meta-analysis on the example of 
the data sets available at $\sqrt{s_{NN}}\,=$ 4.87 GeV. The results and their  interpretation are given in section IV,
while section V contains our conclusions and practical suggestions for planning experiments.

\section{ANALYSIS OF  DATA AND MODELS}

As it was discussed above, the existing approaches \cite{Gazd_rev:10, RafLet} suggest that the onset of deconfinement begins somewhere between the  highest AGS and the lowest SPS energies, i.e. in the collision energy range $\sqrt{s_{NN}}\,=$ 4.2 GeV  \cite{Bugaev:2014a, Bugaev:2014b} --  7.6 GeV \cite{Horn, Step, Gazd_rev:10}. The main reason for such a range is that  almost all irregularities  observed  either at kinetic 
freeze-out  \cite{Horn, Step, Gazd_rev:10}  or
at chemical  freeze-out   \cite{RafLet, Bugaev:2014a, Bugaev:2014b,Bugaev:2013irr} belong to this energy range.  Therefore,  to compare the QGP and HG models we choose the AGS data measured  in Au+Au collisions at 
$\sqrt{s_{NN}}\,=$ 3.1  GeV as the lower bound of collision energy. 
On the other hand,  we extended the upper bound of collision energy from $\sqrt{s_{NN}}\,=$ 7.6 GeV to  $\sqrt{s_{NN}}\,=$17.3 GeV, since there are arguments \cite{Quarkyonic} that at  $\sqrt{s_{NN}}\, \simeq $ 10 GeV there may exist the 
tricritical endpoint of the strongly interacting matter  phase diagram, which in \cite{Quarkyonic} is mistakenly called a triple point. 
Hence, we will be able to investigate the wider energy range than the most interesting one. 

Our choice of collision energies $\sqrt{s_{NN}}\,=$   3.1, 3.6, 4.2, 4.87, 5.4, 6.3, 7.6, 8.8,   12.3 and 17.3 GeV is also dictated by the fact that exactly for these  energies  there were multiple efforts to describe the experimental data  both
by the QGP and by the HG generators. Therefore, in the chosen collision  energy range a  comparison of 
the QDD provided by these two types of models can be made in  details. 
 The collision energies  $\sqrt{s_{NN}}\,\le $  4.87 GeV correspond to  Au+Au  reactions studied at AGS. At 
 $\sqrt{s_{NN}}\,=$ 5.4 GeV the reactions Pb+Si, Si+Si and Si+Al were also studied at AGS, 
 while higher values of collision energy correspond to Pb+Pb reactions investigated at SPS. 
 
 The main object of our meta-analysis is the mean deviation squared  of the quantity $A^{model,h}$ of the model  M from the data $A^{data,h}$ per number of  the data points $n_d$ for  a given particle type $h$  
 \begin{eqnarray}\label{EqI}
 \langle\chi^2/n\rangle^h_A\biggl|_{M} = \frac{1}{n_d} \sum\limits_{k=1}^{n_d} \left[ \frac{A^{data,h}_k -A^{model,h}_k }{\delta A_k^{data,h}}\right]^2  \biggl|_{M}  \,,
\end{eqnarray}
where $\delta A_k^{data,h}$ is an experimental error of the experimental quantity $A^{data,h}_k$ and the summation in Eq. (\ref{EqI})  runs over all data point at given collision energy. To get  the most complete  picture of the  A+A collision process dynamics, we   have to compare the available data on 
  the transverse mass ($m_T$) distributions $A= \frac{1}{m_T}\,\frac{d^2 N(m_T, y)}{d m_T dy}$, the longitudinal rapidity ($y$) distributions $A=\frac{d N(y)}{dy}$ and the hadronic yields (Y) measured  at  midrapidity $A=\frac{d N(y=0)}{dy}$  or/and the total one, i.e. measured within   4$\pi$ solid angle, 
since right  these observables are traditionally believed to be sensitive to the equation of state properties \cite{Shuryak}.

Based on the theory of measurements \cite{Taylor82}, we consider the set of  quantities $\langle\chi^2/n\rangle^h_A\biggl|_{M}$ as the results of the meta-measurements of  the same 
meta-quantity, i.e. the  QDD of  $\{ A^{data,h}_k \}$ data, with the different meta-devices, i.e. models $\{M\}$, with the hadronic probe $h$.
   The  whole set of  quantities $\langle\chi^2/n\rangle^h_A\biggl|_{M}$ allows one to find the mean value $\langle\chi^2/n\rangle_{M}$ properly averaged   over the experimental data and over models belonging to the same class, i.e. M=HG or M=QGP, and over all hadronic species.  Using the averaged values for two classes of models $\langle\chi^2/n\rangle_{HG}$ and $\langle\chi^2/n\rangle_{QGP}$, we could find the regions of their  preferential  and their comparable  description of the experiments. The latter case could provide us with the most probable collision energy range of the hadron-QGP mixed phase threshold. 
 Unfortunately, the published articles usually do not provide one with the quantities $\langle\chi^2/n\rangle^h_A\biggl|_{M}$.
 An exception is the article \cite{S4.9a}, in which one can find the desired values for the $m_T$ spectra of $\Lambda$-hyperons  for 4 rapidity intervals measured in Au+Au collisions at highest AGS energy. Therefore,  using the modern software we calculated the set of  quantities $\langle\chi^2/n\rangle^h_A\biggl|_{M}$ and their errors $\delta \langle\chi^2/n\rangle^h_A\biggl|_{M}$ from the published papers. 
 
 The theoretical models  taken  for  the present analysis belong to two groups:
\begin{itemize}

\item {\bf The used HG models} are as follows: ARC \cite{Arc},  RQMD2.1(2.3) \cite{RQMD}, HSD \cite{HSDgen,S5.4a},  UrQMD1.3(2.0, 2.1, 2.3) \cite{Bass98}, statistical hadronization model (SHM) \cite{SHMgen} and AGSHIJET\_N* \cite{HIJETa,S5.4b}.  These models  do not include the QGP formation in the process of  A+A collisions.
The results of the HG models were taken from the following publications: ARC \cite{S4.9a, S5.4b},  RQMD2.1(2.3) \cite{S4.9a,S4.9b}, HSD \cite{S5.4a,S4.9KL},  UrQMD1.3(2.0, 2.1, 2.3) \cite{S4.9KL, Blume}, SHM \cite{SHM, Blume} and AGSHIJET\_N* \cite{S5.4b}. 
Further  details on what data at what energies were analyzed are presented  in  Tables I and III-XI.

\item {\bf The used QGP models} are as follows:  Quark Combination  (QuarkComb) model \cite{QComb}, 3-fluid dynamics (3FD) model \cite{3FDgen,3FDII,3FDIII},  PHSD model \cite{PHSDgen,Phsd} and Core-Corona model \cite{CoreCorMa,CoreCorMb}.   
These  generators   explicitly assume the QGP formation in A+A collisions.
The results of the  QGP models were taken from the following publications: 
QuarkComb model  \cite{QComb}, 3FD model \cite{3FDII,3FDIII},  PHSD  model \cite{Phsd} and Core-Corona model \cite{CoreCor}.
More details on the  analyzed  data and energies can be found in   Tables I and III-XI.
\end{itemize}
{A short description of these models along with the criteria of their selection can be found in the Appendix.}
%\vspace*{77mm}
%%%\newpage

Our meta-analysis requires  the well-defined errors for the quantities $ \langle\chi^2/n\rangle^h_A\biggl|_{M}$. 
They were defined according to the rule of indirect measurements \cite{Taylor82} as 
 \begin{eqnarray}\label{EqII}
 \Delta_{A}\langle\chi^2/n\rangle^h_A\biggl|_{M} &\equiv& \left[ \sum\limits_{k=1}^{n_d}  \left[\delta A_k^{data,h} \,  \frac{\partial \langle\chi^2/n\rangle^h_A\biggl|_{M}}{\partial ~A^{data,h}_k} \right]^2  \right]^\frac{1}{2}   
 \nonumber \\
&=&
%%\hspace*{-2.2mm}=
 \frac{2}{n_d} \left[  \sum\limits_{k=1}^{n_d}  \left[ \frac{A^{data,h}_k -A^{model,h}_k }{\delta A_k^{data,h}}\right]^2 \biggl|_{M}\right]^\frac{1}{2} \equiv \frac{2}{\sqrt{n_d}} \sqrt{\langle\chi^2/n\rangle^h_A\biggl|_{M}} \,,~
\end{eqnarray}
where in deriving the second equality above we calculated the partial derivatives using Eq. (\ref{EqI}) and then applied Eq. (\ref{EqI}) once more. 

In order to  thoroughly estimate a correspondence between the experimental data and their model description
it is necessary to have  very detailed experimental data which cover rather wide kinematic region and include many 
hadronic species. In practice, however,  the available experimental information is rather limited and, additionally, 
its comparison  with theoretical models in many cases is done not for all available data, but for certain sets only. 
Therefore, first of all we restricted our probes  to the strange particles which include charged kaons $K^\pm$, $K^0_s$ and $\phi$ mesons, and also  $\Lambda (+\Sigma^0)$, $\bar \Lambda$, $\Xi^\pm$ and $\Omega^\pm$ hyperons.  This choice was dictated by the fact that strange particles are the ``clean'' probes, since they are created  at primary/hard collisions. As it was mentioned in  the Introduction, several existing ``signals'' of deconfinement transition are based on the characteristics of $K$-mesons \cite{Horn, Step, Gazd_rev:10}, hence it was natural to consider the strange particles first.

Then for a given probe $h$ and an  observable  $A$ we calculated the average of $ \langle\chi^2/n\rangle^h_A\biggl|_{M}$ over the models of the same class as 
 \begin{eqnarray}\label{EqIII}
 \langle\chi^2/n\rangle^h_A\biggl|_{\omega \overline{M}}  =  \sum\limits_{M=1}^{N_M} \omega (M) \, \langle\chi^2/n\rangle^h_A  \biggl|_{M}  \,,
\end{eqnarray}
where the symbol ${\overline{M}}$ defines the class of models, i.e. ${\overline{M}}  \in \{ HG; QGP\}$,  which are averaged with the weights $\omega (M)$. Here $N_M$ is the number of used theoretical models. 

In order to verify the stability of our findings we employed two 
ways of averaging in  (\ref{EqIII}). First of them  is an arithmetic averaging  with the equal weight for all models, i.e. 
 \begin{eqnarray}\label{EqIV}
 \omega (M| aa) = \frac{1}{N_M} \,, 
\end{eqnarray}
where  $N_M$ is the upper limit of the sum in (\ref{EqIII}), or the number of  used models. For the arithmetic averaging we calculated the averaged  errors  according to the formula 
 \begin{eqnarray}\label{EqV}
 \Delta_{A}\langle\chi^2/n\rangle^h_A\biggl|_{aa\overline{M}}  = \frac{1}{N_M}   \left[  \sum\limits_{M}^{N_M} \left[  \Delta_{A}\langle\chi^2/n\rangle^h_A\biggl|_{M} \right]^2 \right]^\frac{1}{2} \,,
\end{eqnarray}
which is similar to the definition (\ref{EqII}) of the $\langle\chi^2/n\rangle^h_A\biggl|_{aa\overline{M}}$ error.

Besides,  we used another weighted averaging with the weights defined via the errors (\ref{EqII}) of each model   as
 \begin{eqnarray}\label{EqVI}
 \omega (M| wa) = \frac{1}{\left[  \Delta_{A}\langle\chi^2/n\rangle^h_A\biggl|_{M} \right]^2}  \,  \frac{1}{ \sum\limits_{M}^{N_M} \left[  \Delta_{A}\langle\chi^2/n\rangle^h_A\biggl|_{M} \right]^{-2} } \,. 
\end{eqnarray}
For  the weighted averaging (\ref{EqVI}) the best estimate for the average error is given by \cite{Taylor82}
 \begin{eqnarray}\label{EqVII}
 \Delta_{A}\langle\chi^2/n\rangle^h_A\biggl|_{wa\overline{M}}  =   \frac{1}{ \left[   \sum\limits_{M}^{N_M} \left[  \Delta_{A}\langle\chi^2/n\rangle^h_A\biggl|_{M} \right]^{-2} \right]^\frac{1}{2} } \,.
\end{eqnarray}
This kind of averaging is used, if there exist $N_M$  separate measurements $\langle\chi^2/n\rangle^h_A\biggl|_{M1}, \langle\chi^2/n\rangle^h_A\biggl|_{M2}$,... of the same quantity and if the corresponding errors $\Delta_{A}\langle\chi^2/n\rangle^h_A\biggl|_{M1},  \Delta_{A}\langle\chi^2/n\rangle^h_A\biggl|_{M2}$,... are known as well \cite{Taylor82}. In our case the 
measured quantity is the  quality  description of  the observable $A$.

Each way of averaging has its own advantages. Thus, the arithmetic averaging with the weights (\ref{EqIV}) includes all measurements  on equal footing and, hence, as we will see, it allows one to equally account for the contributions  coming from different kinematic regions. 
On the other hand, the weighted averaging of Eq. (\ref{EqVI}) `prefers' the measurements with the smallest  value of the  
QDD  error   and it provides the best estimate for the measured quantity  
\cite{Taylor82}, i.e. for  QDD in our case.
{Note that we also used alternative ways to average   the QDD,  but usually found the results similar to one of two ways of averaging used here. Therefore,   we  concentrate on the averaging methods  given by  Eqs.  (\ref{EqIV}) and (\ref{EqVI}), since they are more convenient than the  other ones and they have a well-defined meaning within the theory of measurements \cite{Taylor82}.
}

These two ways of averaging are used further on for averaging over the measurable  quantities $A =\{\frac{1}{m_T}\,\frac{d^2 N(m_T, y)}{d m_T dy}; \frac{d N(y)}{dy}; Y \}$
 \begin{eqnarray}\label{EqVIII}
 \langle\chi^2/n\rangle^h_{\omega \overline{A}}\biggl|_{\omega \overline{M}} & = & \sum\limits_{A}^{N_A} \omega (A) \, \langle\chi^2/n\rangle^h_A  \biggl|_{\omega \overline{M}}  \,, \quad {\rm with} \\
 \label{EqIX}
 \omega (A| aa) &=& \frac{1}{N_A} \quad  {\rm for ~the~arithmetic~averaging}\,, \\
  \label{EqX}
  \omega (A| wa) &=& \frac{\left[  \Delta_{A}\langle\chi^2/n\rangle^h_A\biggl|_{wa\overline{M}} \right]^{-2}}{ \sum\limits_{N_A}^{A} \left[  \Delta_{A}\langle\chi^2/n\rangle^h_A\biggl|_{wa\overline{M}} \right]^{-2} }  \quad  {\rm for ~the~weighted~averaging} \,. 
\end{eqnarray}
Corresponding errors are calculated using  expressions similar to Eq. (\ref{EqV}) for the arithmetic averaging and
to Eq. (\ref{EqVII}) for the weighted averaging. Then we used these two ways of averaging to calculate the mean values 
$ \langle\chi^2/n\rangle^h_{\omega \overline{A}}\biggl|_{\omega \overline{M}}$ over the hadronic species $h$, but   we do not mix the ways of averaging with each other.  

It is necessary to mention that in some cases  before the averaging over the measurable  quantities $A$ it was necessary to average several sets of the data existing  for the same quantity $A$.  For example,  for  $\sqrt{s_{NN}}\,=$  4.87 GeV 
the yields of charged kaons are known  for   midrapidity and  in full 4 $\pi$ solid angle. Corresponding 
values $\langle\chi^2/n\rangle^h_Y  \biggl|_{M}$ were found first for three HG models (HSD, UrQMD1.3 and UrQMD2.1). 
Then they were averaged over two sets of yields (at midrapidity and the full one), and only after these steps they were averaged over  the types of kaons. This sequence can be found from Table I for the  $K^\pm$ set 1    measured at  
the collision energy $\sqrt{s_{NN}}\,=$  4.87 GeV.

{
Similarly, we performed averaging, if for the same quantity $A$ there were available  data  in different kinematic regions. 
For instance,  for  the collision energy $\sqrt{s_{NN}}\,=$  4.87 GeV the RQMD2.1 model \cite{S4.9a} provides the description of  $m_T$ distributions  of $\Lambda$ hyperons for  four  intervals of longitudinal rapidity in the range $2 < y < 3.2$  (see Fig. 2  \cite{S4.9a} and $\Lambda$ set 1 in Table I). On the other   hand,   two versions of RQMD2.3 model were  used in \cite{S4.9b} to describe  the  $m_T$ distributions  of  $\Lambda$ hyperons for other four longitudinal rapidity in the range $2.2 < y < 3.4$
(see  Fig. 5   in \cite{S4.9b}). 
Therefore, first of all  it was necessary  to determine   the QDD  of  $\Lambda$ hyperons  
over $m_T$ distributions for   these models  at each rapidity interval $y_k$ and then to average  the obtained  values 
over all rapidity intervals.  Then the  $m_T$ distribution results  of  $\Lambda$ hyperons  found for two versions of RQMD2.3 model \cite{S4.9b} were further averaged with the QDD for the longitudinal rapidity distributions
(see  Fig. 7   in \cite{S4.9b} and $\Lambda$ set 2 in Table I).
Such  information and the final results can be found   in Tables I and III-XI  for the arithmetic averaging. For the weighted averaging (\ref{EqVII})  the subsequence of steps was absolutely similar and, hence, for such a way of averaging we give  the final results only.  
In the next section we demonstrate  a detailed way of finding  the QDD for the collision energy  
$\sqrt{s_{NN}}\,=$  4.87 GeV.
}

\section{Details of averaging procedure  for $\sqrt{s_{NN}}\,=$  4.87 GeV}

\begin{figure}[h]
\centerline{\includegraphics[width=112mm]{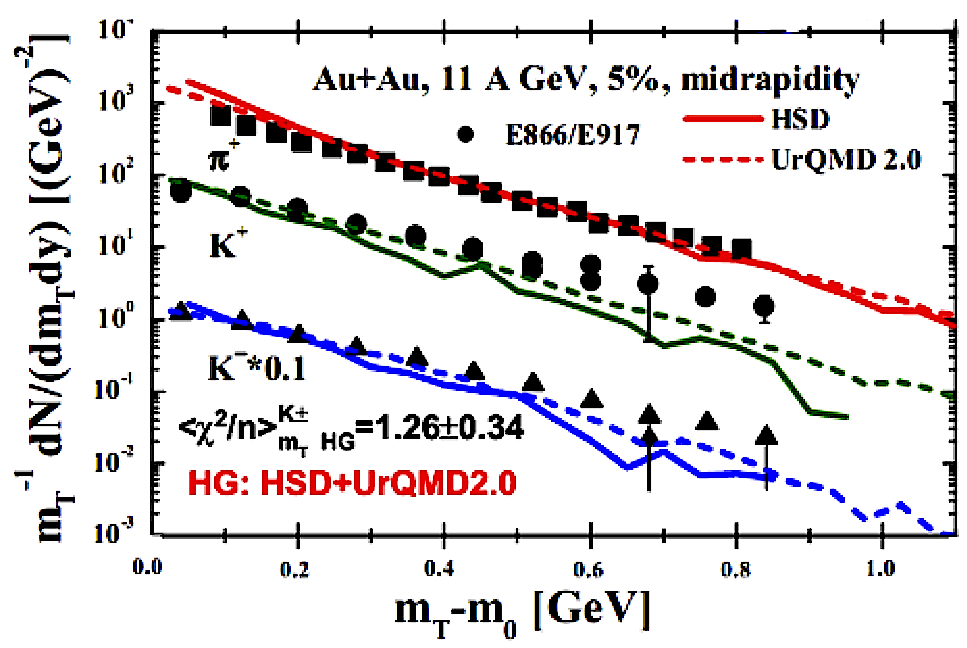}}
\caption{(Color online) Comparison of transverse mass spectra of $K^+$ and $K^-$ mesons at midrapidity from HSD (solid curves) and UrQMD 2.0 (dashed curves) for  $\sqrt{s_{NN}}\,=$  4.87 GeV taken from Fig. 7 of  \cite{S4.9KL}.
The result for  $ \langle\chi^2/n\rangle^{K^\pm}_{m_T}\biggl|_{aa\overline{HG}}$ shown in this figure is described in the text.
}
\label{fig1A}
\end{figure}

To explain in more details the procedure of averaging, in this section we consider it on the example 
of  our meta-analysis for the collision energy $\sqrt{s_{NN}}\,=$  4.87 GeV, i.e. for laboratory energy $E_{lab} = 10.7$ GeV.  Let us for this purpose analyze the 
$K^\pm$ set 1 from the Table I.  To get the QDD  of  $m_T$ spectra  $ \langle\chi^2/n\rangle^{K^\pm}_{m_T}\biggl|_{aa\overline{HG}}$ for these mesons   we used  Fig. 7 from \cite{S4.9KL}. It is also shown here as Fig.~\ref{fig1A}.
\begin{table}[t]
\begin{tabular}{|c|c|c|c|} \hline                   
                     &  \multicolumn{3}{c|}{$\sqrt{s_{NN}}\,=$  4.87 GeV}      \\ \hline
                                       & $m_{T}$-distribution & rapidity distribution & Yields       \\ \hline
$\langle\chi^2/n\rangle =$     &  $1.26\pm0.34$  & $2.353\pm0.626$  &  $4.3\pm1.2$ $\left(\frac{d N}{d y}\bigl|_{y=0} \& \quad4 \pi\right)$ \\ 
$K^\pm$   set 1    & HSD \& UrQMD2.0 & QuarkComb. model&   HSD \& UrQMD1.3(2.1)  \\ 
                   &  Fig.7,  Ref. \cite{S4.9KL}& Fig.5 Ref. \cite{QComb} &  Fig.1, 2 Ref. \cite{S4.9KL}     \\ \hline 
$\langle\chi^2/n\rangle = $    & $ 1.23\pm0.22$&                    &   \\ 
$K^\pm$   set 2    &3 versions of HSD \& UrQMD2.1 & N/A &  N/A \\ 
                   & Figs. 8, 10, 12  Ref. \cite{S4.9KL} &   & \\ \hline
$\langle\chi^2/n\rangle =$     & $1.15\pm0.65$ &                    &  $7.65\pm5.53$   \\ 
$K^{+}$             & 3FD           &          N/A &  3FD     \\
                   & Fig.1, Ref. \cite{3FDIII}&   & Fig.9, Ref. \cite{3FDII} \\ \hline
$\langle\chi^2/n\rangle =$     & $1.51\pm0.74$ &                    &  $0.15\pm0.775$   \\
$K^{-}$              & 3FD           &    N/A   &  3FD     \\
                    & Fig.1, Ref. \cite{3FDIII}&   & Fig.9, Ref. \cite{3FDII} \\ \hline
$\langle\chi^2/n\rangle =$     &$2.54\pm0.01, 1.07\pm0.002$  & $2.75\pm1.66,$ $5.74\pm2.1$  & $2.6\pm1.3$ $\left(\frac{d N}{d y}\bigl|_{y=0} \& \quad4 \pi\right)$ \\
$\Lambda$ set 1    & ARC,RQMD2.1   &   ARC,RQMD2.1                &   HSD \& UrQMD1.3(2.1)  \\
                   &Fig. 2 Ref.  \cite{S4.9a}& Fig. 4 Ref.  \cite{S4.9a} &  Fig. 1 Ref. \cite{S4.9KL}     \\ \hline 
$\langle\chi^2/n\rangle =$     &  $3.65\pm0.6,$ $2.4\pm0.55$ & $4.67\pm1.155$     &      \\ 
 $\Lambda$ set 2   & $m_T$+y:RQMD2.3(cascade), & QuarkComb. model  &   N/A \\ 
                   &  RQMD2.3(mean-field)       &                   &  \\                             
                   &  Figs. 5, 7  Ref. \cite{S4.9b}      & Fig. 5  Ref. \cite{QComb}     & \\ \hline 
$\langle\chi^2/n\rangle =$     &               &                   & $3.46\pm3.72,$ $3.01\pm3.5$ \\
$\phi$             &        N/A         &           N/A          & SHM, UrQMD       \\
                   &               &                   & Fig. 17  Ref. \cite{Blume} \\  \hline
                   
\end{tabular}
\caption{The QDD  provided by  HG and QGP  models. The 1-st column indicates the particle species,
the 2-nd one  shows results for  the quality of $m_T$ spectra description,  the 3-rd one shows results
for the quality of $y_L$ spectra   description, while the 4-th one gives  results for  the  QDD  of yields 
 at the collision energy $\sqrt{s_{NN}}\,=$  4.87 GeV.  In some  rows  there are   two values of  $\langle\chi^2/n\rangle $ which correspond to the models and references shown below in the same column. For more details see the text.
}
\label{tableI}
\end{table}

The scan of curves and experimental data points allowed us to get the following results for the arithmetic averaging
\begin{eqnarray}\label{EqXI}
&& \left. \begin{array}{l}
\langle\chi^2/n\rangle^{K^+}_{m_T}\biggl|_{HSD} \hspace*{-2.2mm}= 1.9 \pm 0.83  \\
\langle\chi^2/n\rangle^{K^+}_{m_T}\biggl|_{UrQMD2.0} \hspace*{-4.4mm}= 1.05 \pm 0.62 \\
\end{array} \right\} ~\Rightarrow~\langle\chi^2/n\rangle^{K^+}_{m_T}\biggl|_{aa\overline{HG}} \hspace*{-4.4mm}= 
1.47 \pm 0.52 \,,\\
&& \left. \begin{array}{l}
\langle\chi^2/n\rangle^{K^-}_{m_T}\biggl|_{HSD} \hspace*{-2.2mm}= 1.35 \pm 0.7 \\
\langle\chi^2/n\rangle^{K^-}_{m_T}\biggl|_{UrQMD2.0} \hspace*{-4.4mm}= 0.75 \pm 0.52\\
\end{array} \right\} ~\Rightarrow~\langle\chi^2/n\rangle^{K^-}_{m_T}\biggl|_{aa\overline{HG}} \hspace*{-4.4mm}= 
1.05 \pm 0.44 \,.
\label{EqXII}
\end{eqnarray}
Averaging the above results over types of hadrons one finds that $ \langle\chi^2/n\rangle^{K^\pm}_{m_T}\biggl|_{aa\overline{HG}} = 1.26 \pm 0.34$. At the same time the weighted averaging gives us practically the same result $ \langle\chi^2/n\rangle^{K^\pm}_{m_T}\biggl|_{wa\overline{HG}} = 1.125 \pm 0.32$.

\begin{figure}[h]
\centerline{\includegraphics[width=84mm]{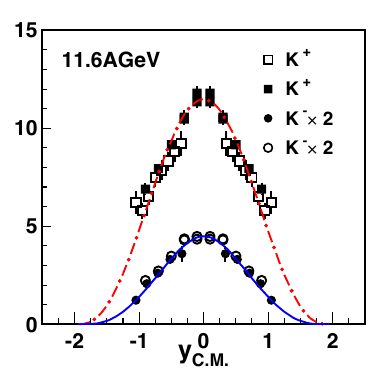}}
\caption{(Color online) The longitudinal rapidity spectra  of $K^+$ and $K^-$ mesons (symbols) reproduced by the QuarkComb.  model \cite{QComb}.
This is part of Fig. 5 of   \cite{QComb} which was used to determine $ \langle\chi^2/n\rangle^{K^\pm}_{y_L}\biggl|_{aa\overline{QGP}}$.
}
\label{fig2A}
\end{figure}

In the same way we determined the QDD of the  longitudinal rapidity  distributions of kaons $ \langle\chi^2/n\rangle^{K^\pm}_{y_L}\biggl|_{aa\overline{QGP}}$ for $\sqrt{s_{NN}}\,=$  4.87 GeV using the results of QuarkComb. model \cite{QComb} (see Fig.~\ref{fig2A} and  $K^\pm$ set 1 in Table I for details)
\begin{eqnarray}
&& \left. \begin{array}{l}
\langle\chi^2/n\rangle^{K^+}_{y_L}\biggl|_{QuarkComb} \hspace*{-2.2mm}= 4.47 \pm 1.22  \\
\langle\chi^2/n\rangle^{K^-}_{y_L}\biggl|_{QuarkComb} \hspace*{-4.4mm}= 0.236 \pm 0.28 \\
\end{array} \right\} ~\Rightarrow~\langle\chi^2/n\rangle^{K^\pm}_{y_L}\biggl|_{{QuarkComb}} \hspace*{-4.4mm}= 
2.353 \pm 0.63 \,.
\label{EqXIII}
\end{eqnarray}
The weighted averaging gives us a different result $\langle\chi^2/n\rangle^{K^\pm}_{y_L}\biggl|_{{QuarkComb}} \hspace*{-4.4mm}= 0.448 \pm 0.273 $, which within error bars is more close to the value of negative kaons in  (\ref{EqXIII}).

\begin{figure}[h]
\centerline{\includegraphics[width=112mm]{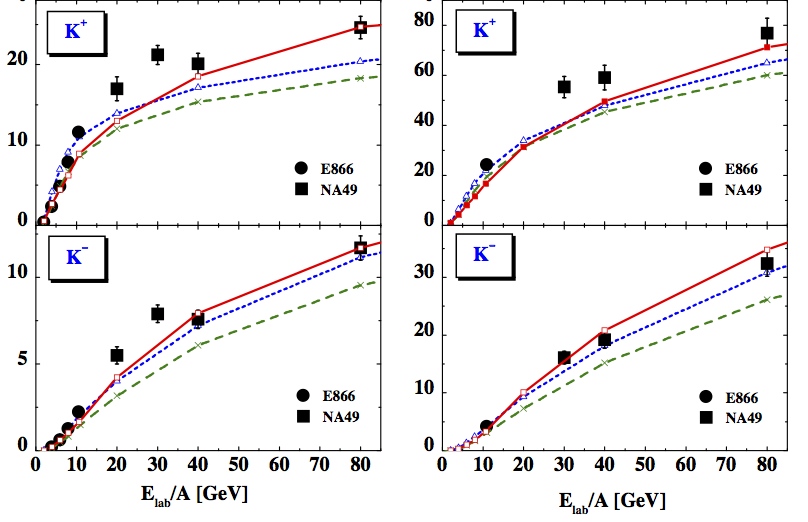}}
\caption{(Color online) The yields of $K^+$ and $K^-$ mesons (symbols) reproduced by the following HG models: 
HSD (solid curves),  UrQMD1.3 (short dashed curve)  and UrQMD2.1 (long dashed curve). The left column corresponds
to the particle yields measured at midrapidity, while the right column gives the total yields in $4\pi$ solid angle.
This is part of Fig. 2 of    \cite{S4.9KL} which was used to determine $ \langle\chi^2/n\rangle^{K^\pm}_{\frac{d N}{d y}}\biggl|_{aa\overline{HG}}$ and  $ \langle\chi^2/n\rangle^{K^\pm}_{4\pi}\biggl|_{aa\overline{HG}}$ for  all energies given in the plots.
}
\label{fig3A}
\end{figure}

Similarly, from Fig. 2 of   \cite{S4.9KL} we found the QDD  of  the midrapidity  multiplicity of kaons  for $\sqrt{s_{NN}}\,=$  4.87 GeV, i.e. $ \langle\chi^2/n\rangle^{K^\pm}_{\frac{d N}{d y}}\biggl|_{aa\overline{HG}}$. In particular, from Fig.~\ref{fig3A}
we determined the desired quantities for all values of collision energy given in this figure
\begin{eqnarray}
\label{EqXIV}
&& \left. \begin{array}{l}
\langle\chi^2/n\rangle^{K^+}_{\frac{d N}{d y}}\biggl|_{HSD} \hspace*{-2.2mm}= 12.44 \pm 7.05 \\
\langle\chi^2/n\rangle^{K^+}_{\frac{d N}{d y}}\biggl|_{UrQMD1.3} \hspace*{-4.4mm}= 0.32 \pm 1.14\\
\langle\chi^2/n\rangle^{K^+}_{\frac{d N}{d y}}\biggl|_{UrQMD2.0} \hspace*{-4.4mm}= 14.62 \pm 7.64\\
\end{array} \right\} ~\Rightarrow~\langle\chi^2/n\rangle^{K^+}_{\frac{d N}{d y}}\biggl|_{aa\overline{HG}} \hspace*{-4.4mm}= 9.126 \pm 3.5 \,, \\
&& \left. \begin{array}{l}
\langle\chi^2/n\rangle^{K^-}_{\frac{d N}{d y}}\biggl|_{HSD} \hspace*{-2.2mm}= 2.4 \pm 3.1 \\
\langle\chi^2/n\rangle^{K^-}_{\frac{d N}{d y}}\biggl|_{UrQMD1.3} \hspace*{-4.4mm}= 0.4 \pm 1.27\\
\langle\chi^2/n\rangle^{K^-}_{\frac{d N}{d y}}\biggl|_{UrQMD2.0} \hspace*{-4.4mm}= 4.76 \pm 4.36\\
\end{array} \right\} ~\Rightarrow~\langle\chi^2/n\rangle^{K^+}_{\frac{d N}{d y}}\biggl|_{aa\overline{HG}} \hspace*{-4.4mm}= 2.52 \pm 1.83\,.
\label{EqXV}
\end{eqnarray}

In the same way we determined the arithmetic average of the QDD  of the  total $K^\pm$  multiplicities
$ \langle\chi^2/n\rangle^{K^\pm}_{4\pi}\biggl|_{aa\overline{HG}}$
\begin{eqnarray}\label{EqXVI}
&& \left. \begin{array}{l}
\langle\chi^2/n\rangle^{K^+}_{4\pi}\biggl|_{HSD} \hspace*{-2.2mm}= 9.55 \pm 6.2 \\
\langle\chi^2/n\rangle^{K^+}_{4\pi}\biggl|_{UrQMD1.3} \hspace*{-4.4mm}= 1. \pm 2.\\
\langle\chi^2/n\rangle^{K^+}_{4\pi}\biggl|_{UrQMD2.0} \hspace*{-4.4mm}= 4. \pm 4.\\
\end{array} \right\} ~\Rightarrow~\langle\chi^2/n\rangle^{K^+}_{4\pi}\biggl|_{aa\overline{HG}} \hspace*{-4.4mm}= 
4.85 \pm 2.55  \,, \\
&& \left. \begin{array}{l}
\langle\chi^2/n\rangle^{K^-}_{4\pi}\biggl|_{HSD} \hspace*{-2.2mm}= 0.67 \pm 1.63 \\
\langle\chi^2/n\rangle^{K^-}_{4\pi}\biggl|_{UrQMD1.3} \hspace*{-4.4mm}= 0.07 \pm 0.54\\
\langle\chi^2/n\rangle^{K^-}_{4\pi}\biggl|_{UrQMD2.0} \hspace*{-4.4mm}= 1.2 \pm 2.2\\
\end{array} \right\} ~\Rightarrow~\langle\chi^2/n\rangle^{K^-}_{4\pi}\biggl|_{aa\overline{HG}} \hspace*{-4.4mm}
= 0.646 \pm 0.93 \,.
\label{EqXVII}
\end{eqnarray}
From Eqs. (\ref{EqXIV}) and (\ref{EqXVI}) we found the corresponding average  over two multiplicity sets for positive kaons, while Eqs.  (\ref{EqXV}) and (\ref{EqXVII}) allowed us to determine a similar average for negative kaons
\begin{eqnarray}
&& \left. \begin{array}{l}
\langle\chi^2/n\rangle^{K^+}_{aa\overline{\frac{d N}{d y}}}\biggl|_{aa\overline{HG}} \hspace*{-2.2mm}= 7. \pm 2.165  \\
\langle\chi^2/n\rangle^{K^-}_{aa\overline{\frac{d N}{d y}}}\biggl|_{aa\overline{HG}} \hspace*{-4.4mm}= 1.6 \pm 1.026 \\
\end{array} \right\} ~\Rightarrow~\langle\chi^2/n\rangle^{K^\pm}_{aa\overline{\frac{d N}{d y}}}\biggl|_{aa\overline{HG}} \hspace*{-4.4mm}= 
4.3 \pm 1.2\,,
\label{EqXVIII}
\end{eqnarray}
where in the last step we averaged the results over two kinds of  kaons. Repeating the same sequence of steps for the weighted averaging of Eqs. (\ref{EqVI}) and (\ref{EqVII}), we found $\langle\chi^2/n\rangle^{K^\pm}_{wa\overline{\frac{d N}{d y}}}\biggl|_{wa\overline{HG}} \hspace*{-4.4mm}= 
0.5 \pm 0.41$.

A few additional words should be said about various short hand notations used  in Tables  I and III-XI,
which contain only the results of arithmetic averaging. Tables III-XI are given at the end of this work.
To shorten our remarks inside these  tables  the sign $\&$ is used to  demonstrate the fact that the corresponding 
value of  $\langle\chi^2/n\rangle $ is  averaged  over two sets of data or over the results of models. 
For example, the notation  $\frac{d N}{d y}\bigl|_{y=0} \& \quad4 \pi$  used in Table  I  shows  that the value 
of $\langle\chi^2/n\rangle $  is averaged over the yields measured at midrapidity and in the full solid angle.  Similarly, the notation  `HSD \& UrQMD2.0' means  that we give the   value of $\langle\chi^2/n\rangle $ averaged   over the results of models HSD and UrQMD2.0. 
In Table I, the notation  `$m_T$+y'  made for $\Lambda$ hyperons (set 2) means that we give the value of  $\langle\chi^2/n\rangle $ averaged over  $m_T$- and y-distributions. More explaining remarks can be found in the captions of corresponding 
Tables. 
\begin{figure}[h]
\centerline{\includegraphics[width=178mm]{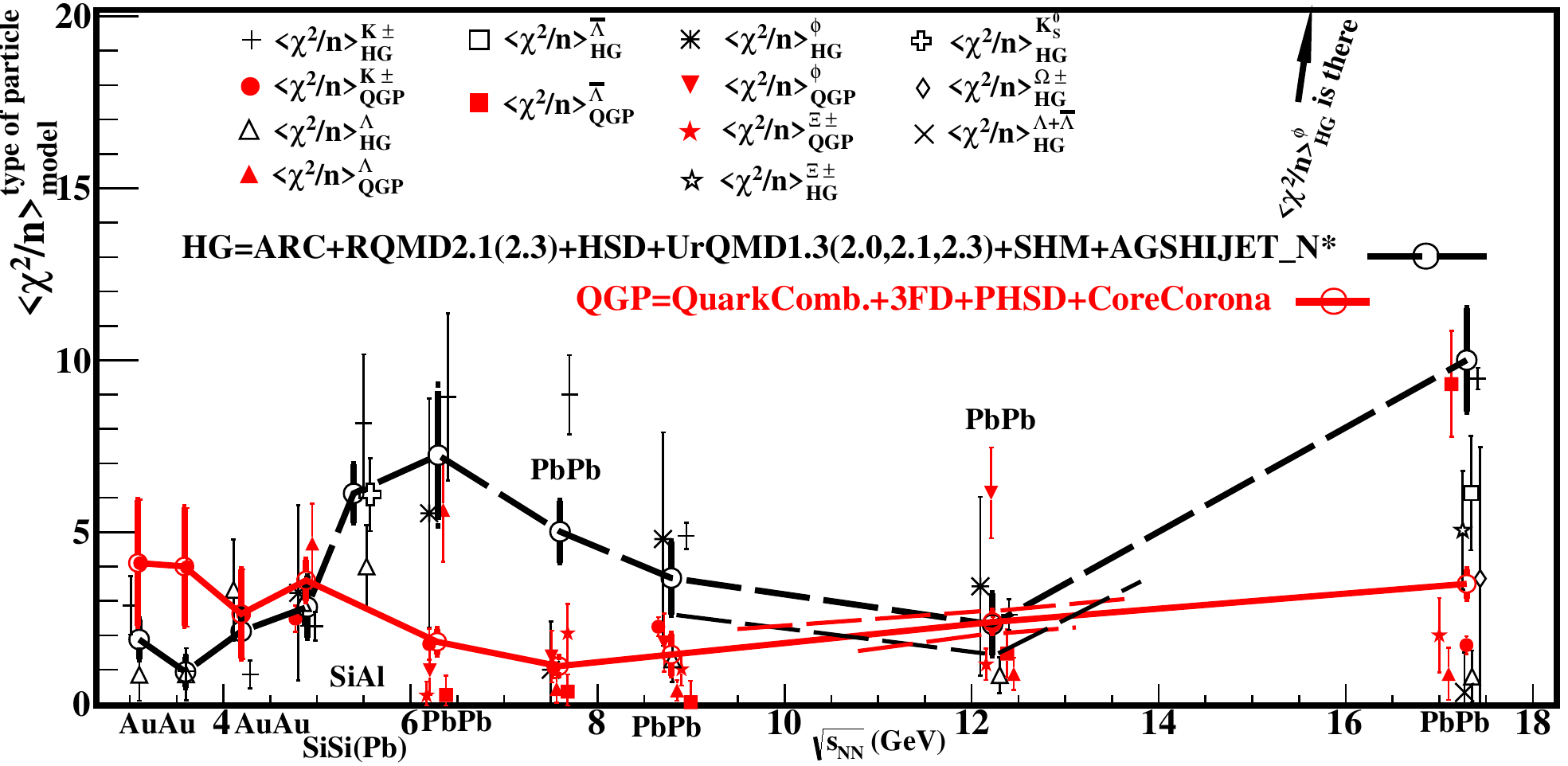}}
\caption{(Color online) KTBO plot 1: Comparison of  $\langle\chi^2/n\rangle^{aa\overline{\{h\}}}_{aa\overline{\{A\}}}\biggl|_{aa\overline{HG}}$
 (black symbols and dashed curve) and 
$\langle\chi^2/n\rangle^{aa\overline{\{h\}}}_{aa\overline{\{A\}}}\biggl|_{aa\overline{QGP}}$  (red symbols and solid curve) as functions of collision energy obtained for the arithmetic averaging. The symbols of different hadrons which  correspond to the same collision energy are slightly spread around the energy value for better perception. 
The symbols are connected by the lines to guide the eye.
}
\label{fig4A}
\end{figure}

\section{RESULTS}

Our major results are shown in Figs.~\ref{fig4A} and \ref{fig6A} and in Table II. The auxiliary results are given in the Tables I, III--XI.  As one can see from the KTBO-plot 1, i.e. Fig.~\ref{fig4A},  there are several matches  of the QDDs  by  HG and QGP  models. At low collision energies the HG quantity $\langle\chi^2/n\rangle^{aa\overline{\{h\}}}_{aa\overline{\{A\}}}\biggl|_{aa\overline{HG}}$ is smaller than the one of QGP models, i.e. QDD  is higher for HG generators. 
{Although at energies below 4.87 GeV  the QGP models are represented by the results of 3FD model, we stress that
it would be extremely hard to reach a better description of the data by these models compared to the HG ones. It is so, since the latter provide almost an excellent description of the analyzed data for  3.1 GeV $ \le \sqrt{s_{NN}} \le 4.2$ GeV, as one can 
see from Table II.
}

At the collision energies $\sqrt{s_{NN}}\,=$  4.2 GeV ,$\sqrt{s_{NN}}\,=$  4.87 GeV and $\sqrt{s_{NN}}\,=$  12.3 GeV one finds 
 \begin{eqnarray}\label{EqXIX}
\langle\chi^2/n\rangle^{aa\overline{\{h\}}}_{aa\overline{\{A\}}}\biggl|_{aa\overline{HG}} & = & \langle\chi^2/n\rangle^{aa\overline{\{h\}}}_{aa\overline{\{A\}}}\biggl|_{aa\overline{QGP}}  \,. 
\end{eqnarray}
In the collision  energy ranges  4.87 GeV $< \sqrt{s_{NN}} <$  12.3 GeV and  12.3 GeV  $< \sqrt{s_{NN}} <$  17.3 GeV the QGP models describe the data essentially better. Therefore, the arithmetic averaging meta-analysis suggests that 
at energies below 4.2 GeV there is  hadron phase, while in the region  4.2 GeV $\le \sqrt{s_{NN}} \le$  4.87 GeV 
there is   hadron-QGP mixed phase, while at higher energies there exists  QGP.  Such a picture is well fit into the recent findings 
of the  generalized shock  adiabat model  \cite{Bugaev:2014a,Bugaev:2014b}. However, the most interesting question is 
how should we interpret  the coincidence of two sets of results at the collision energy   $\sqrt{s_{NN}} =$  12.3 GeV?
\begin{figure}[h]
\centerline{\includegraphics[width=84mm]{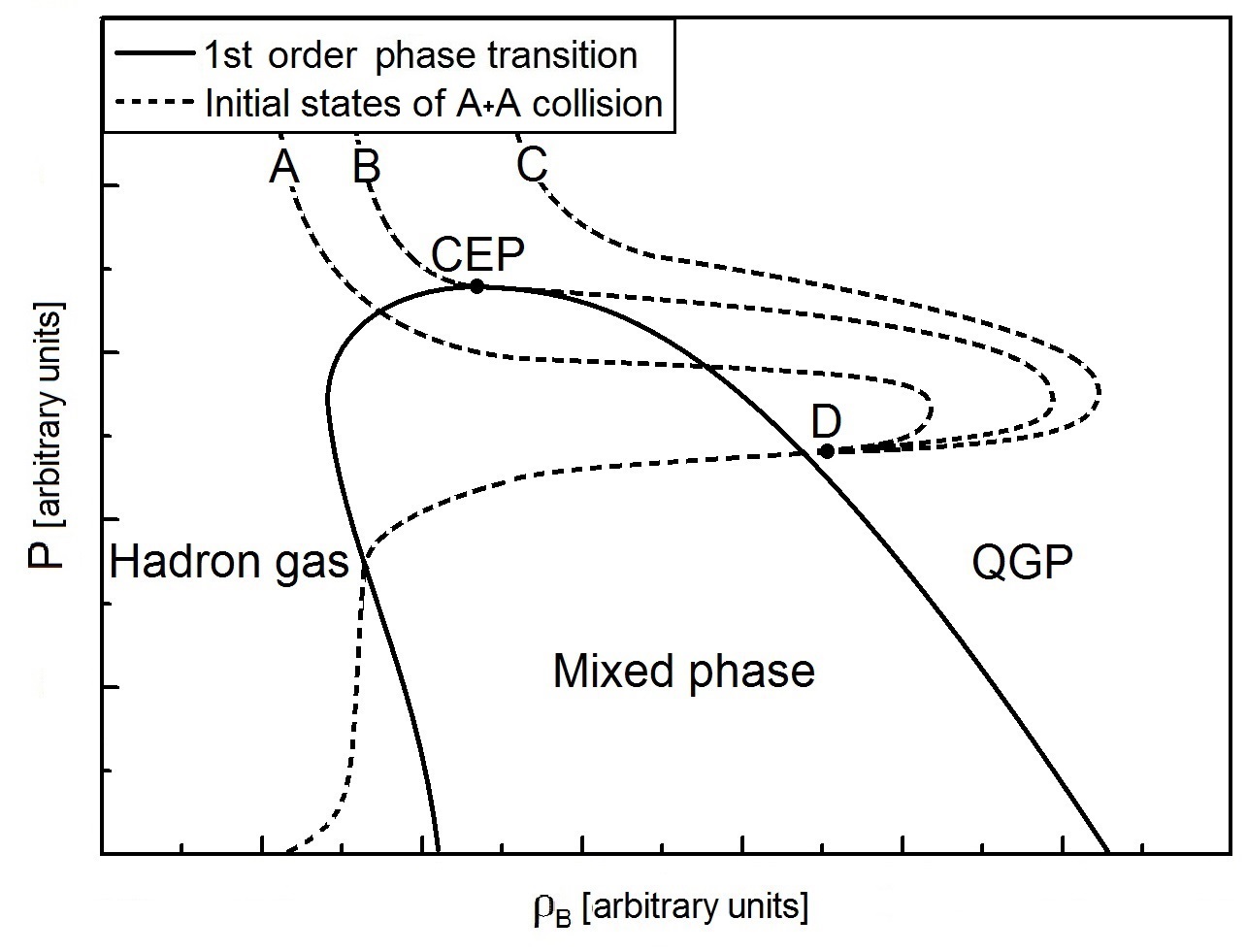} \hspace*{1.1mm}\includegraphics[width=84mm]{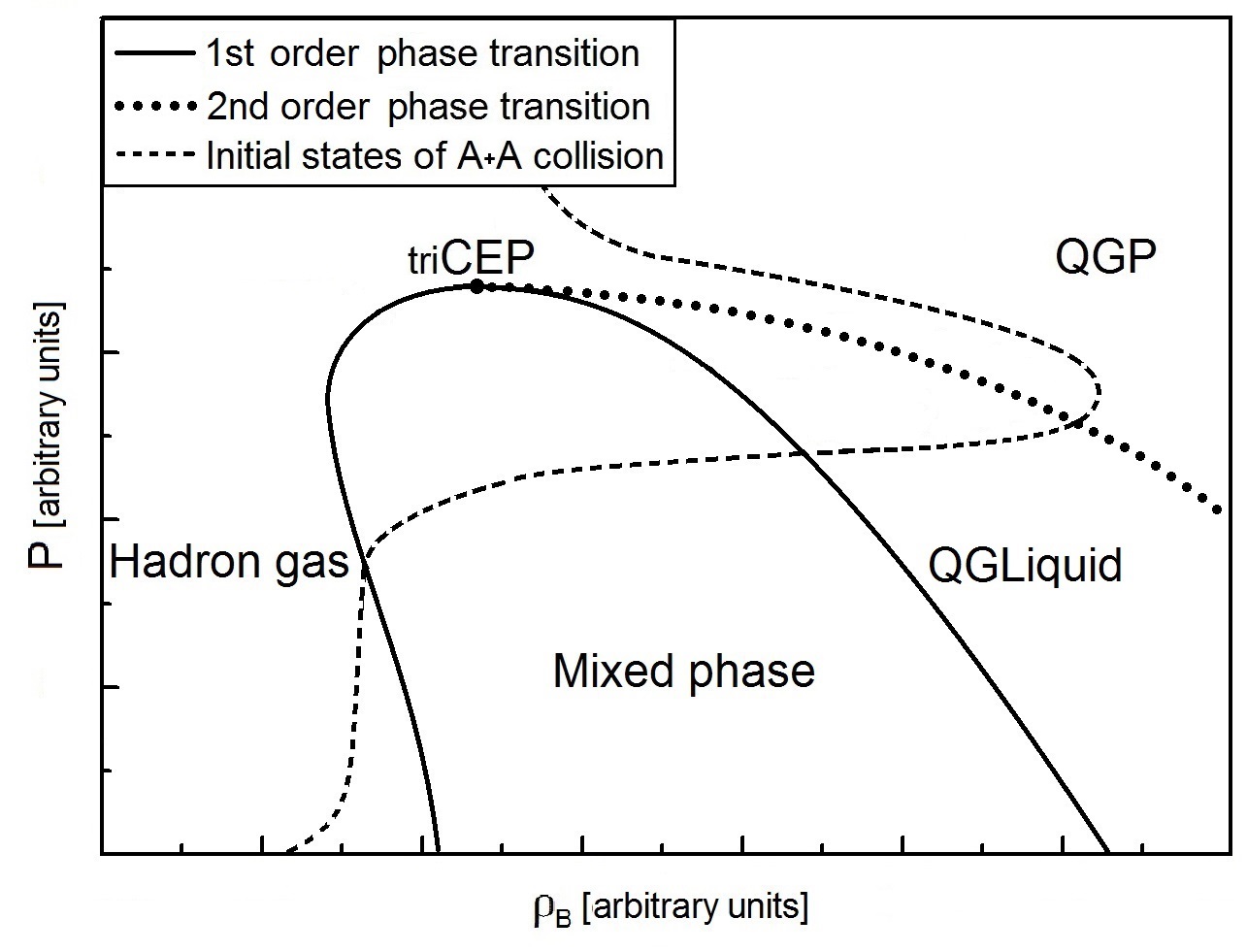}}
\caption{Schematic pictures of  possible locations of the initial states of matter formed  in A+A collisions are shown 
on the plane of baryonic density and pressure.   Each point on these 
trajectories (dashed curves) corresponds to a single collision energy value.  {\bf Left panel:} As it is argued in the text the possible initial states correspond to  the  trajectories AD or BD as it follows from KTBO-plot 1 for the case of critical endpoint.  The trajectory CD is located far from the mixed  phase region and, hence, it cannot generate the second region in which the QDDs of HG  and QGP models are  equally good. 
{\bf Right panel:} In case of the tricritical endpoint the  second region in which the QDDs of HG  and QGP  models are equally good may, alternatively, appear due to the second phase transition. 
}
\label{fig5A}
\end{figure}

At first glance it seems that at the collision energy   $\sqrt{s_{NN}} =$  12.3 GeV the QGP states created by the corresponding generators touch the phase boundary with hadron phase. However,  one must remember that both curves
depicted in  Fig.~\ref{fig4A} have, in fact,  finite width defined by the error bars.  Taking into account an overlap of the  
curves with finite error bars, one immediately concludes that the overlap region is rather wide on collision energy scale, namely it ranges  from $\sqrt{s_{NN}} \simeq$  10 GeV to $\sqrt{s_{NN}} \simeq$  13.5 GeV. Recalling that the collision energy width of the mixed phase at low  values of $\sqrt{s_{NN}}$ is below 1 GeV,  one may guess  that  the most probable scenario is that the initial equilibrated states of the matter formed in A+A collisions at  $\sqrt{s_{NN}} \simeq$  10 GeV 
return into the hadron-QGP mixed phase, while at  $\sqrt{s_{NN}} \simeq$  13.5 GeV they return from the mixed phase into 
QGP.  

From the present meta-analysis results it is hard to tell, whether the initial equilibrated states of the matter formed in A+A collisions at  $\sqrt{s_{NN}} \in $  [10; 13.5] GeV reach the hadronic phase or they correspond to the mixed phase alone. 
However, such a wide range  of collision energy of about 3.8 GeV should  correspond to something entirely new, 
compared to the mixed phase  located at the range $\sqrt{s_{NN}} \in $  [4.2; 4.87] GeV. We believe that the energy range 
$\sqrt{s_{NN}} \in $  [10; 13.5] GeV may correspond to the vicinity of the critical endpoint of the QCD phase diagram. 
The most  probable locations  on phase diagram  of  the  initial states which  formed in A+A collisions are schematically shown in the left panel of Fig.~\ref{fig5A} in case of the critical endpoint. 
Our main reasons in favor of such a hypothesis are as follows. Recalling argumentation of Ref. \cite{Quarkyonic} on the location of (tri)critical endpoint, we have to stress that the vicinity of  collision energy 10 GeV was independently found in this work as the second entrance into the mixed phase from QGP.  On the other hand, the lattice QCD results on the baryonic chemical potential  $\mu_B$ dependence of pseudo-critical temperature $T_{ps} (\mu_B)$ found from maximum of chiral 
susceptibility  \cite{ChirSusc1,ChirSusc2} or  from chiral limit \cite{ChirLim1,ChirLim2}
show that the chemical freeze-out states at $\sqrt{s_{NN}} = $ 17.3 GeV \cite{Bugaev13:Horn,Bugaev13:SFO} 
correspond to the cross-over, but  to the  critical endpoint
 (see \cite{Karsch13} for an extended discussion). 
Therefore, we conclude that such a point should exist at collision energies
below 17.3 GeV, i.e. close to the discussed region of collision energies. 

\begin{figure}[h]
\centerline{\includegraphics[width=178mm]{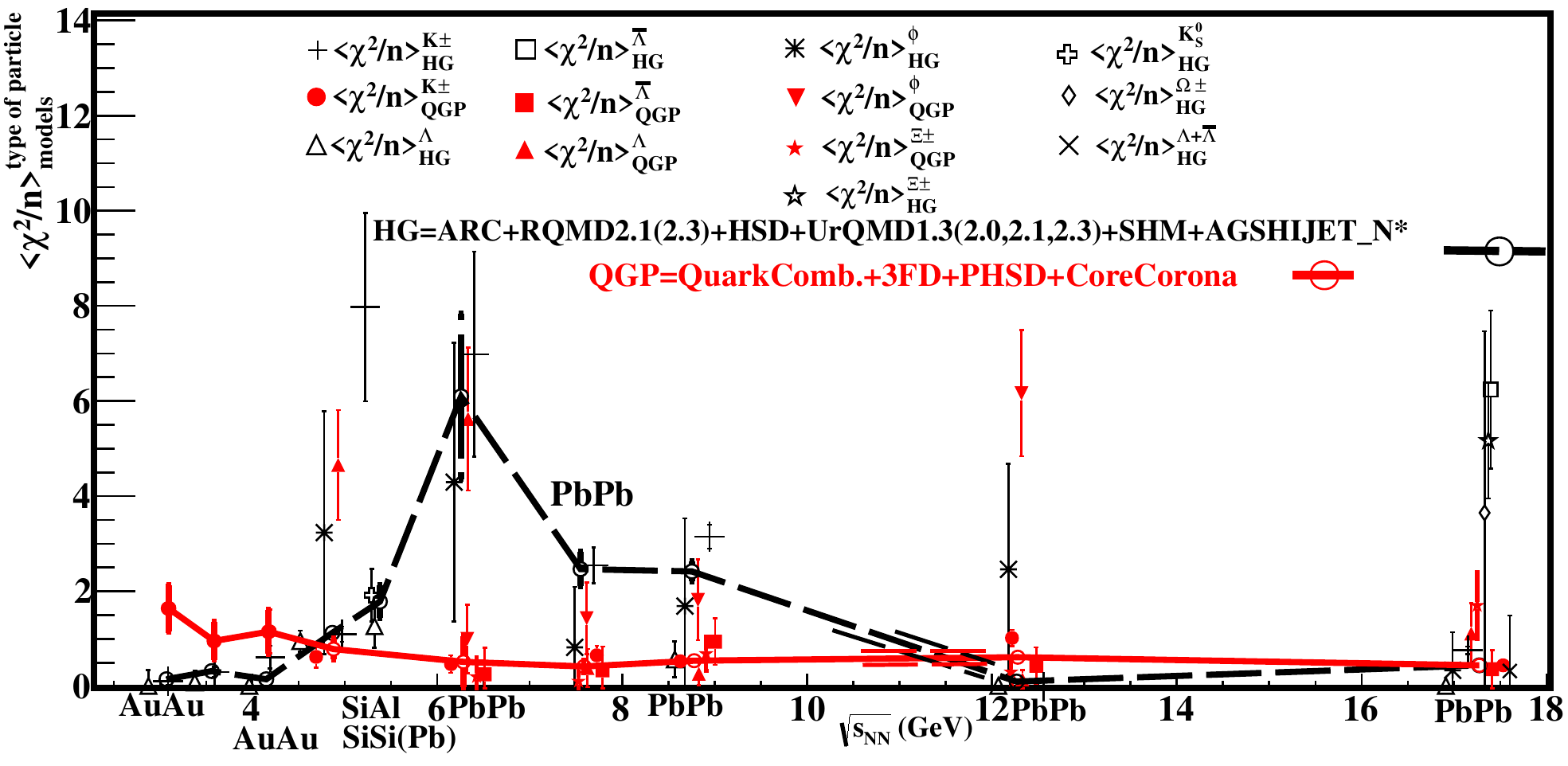}}
\caption{(Color online) KTBO plot 2: Comparison of  $\langle\chi^2/n\rangle^{wa\overline{\{h\}}}_{wa\overline{\{A\}}}\biggl|_{wa\overline{HG}}$
 (black symbols and dashed  curve) and 
$\langle\chi^2/n\rangle^{wa\overline{\{h\}}}_{wa\overline{\{A\}}}\biggl|_{wa\overline{QGP}}$  (red symbols and solid curve) as functions of collision energy obtained for the wighted averaging. The symbols of different hadrons which  correspond to the same collision energy are slightly spread around the energy value for better perception. 
The symbols are connected by the lines to guide the eye. The thin straight lines below collision energy 12 GeV demonstrate
the intersection region due to finite error bars.
}
\label{fig6A}
\end{figure}

Of course, alternatively, the second mixed phase region found here  at $\sqrt{s_{NN}} $= 10--13.5 $=11.75\pm1.75$ GeV  may correspond to the second phase transition (chiral), but even in this case the (tri)critical endpoint of the QCD phase diagram should be located nearby. Such a situation is depicted in the right  panel of Fig.~\ref{fig5A}.

In order to better determine the energy range of the second entrance inside the mixed phase, let us verify our conclusions on the meta-analysis results shown in the KTBO-plot 2 (see Fig.~\ref{fig6A}). This  KTBO-plot contains the meta-analysis results 
for the weighted averaging.  Comparing KTBO-plot 1 and KTBO-plot 2 one can see that in many cases the values of $\langle\chi^2/n\rangle^{wa\overline{\{h\}}}_{wa\overline{\{A\}}}\biggl|_{wa\overline{QGP}}$ and  $\langle\chi^2/n\rangle^{wa\overline{\{h\}}}_{wa\overline{\{A\}}}\biggl|_{wa\overline{HG}}$ are essentially smaller than  the corresponding values obtained by the arithmetic averaging. This is a clear demonstration of the fact that the weighted averaging (\ref{EqVI}) and (\ref{EqVII})
`prefers' the results with minimal error bars, which in our case means the smaller value of  $\langle\chi^2/n\rangle^h_A\biggl|_{M}$
due to Eq.  (\ref{EqII}). {Therefore, one has to be careful with this kind of averaging because at each energy there are only a few  hadronic characteristics which have low value of $\langle\chi^2/n\rangle^h_A\biggl|_{M}$.} Nevertheless, our main conclusions from this plot are qualitatively similar to the case of  the KTBO-plot 1, although there are some quantitative difference. Thus, the equality 
 \begin{eqnarray}\label{EqXX}
\langle\chi^2/n\rangle^{wa\overline{\{h\}}}_{wa\overline{\{A\}}}\biggl|_{wa\overline{HG}} & = & \langle\chi^2/n\rangle^{wa\overline{\{h\}}}_{wa\overline{\{A\}}}\biggl|_{wa\overline{QGP}}  \,,
\end{eqnarray}
holds for the collisions energies $\sqrt{s_{NN}}\,=$  4.87 GeV, $\sqrt{s_{NN}}\,=$10.8--12  GeV and $\sqrt{s_{NN}}\,=$  17.3 GeV. Accounting for the finite error bars, we found that within the error bars the HG and QGP models describe the data equally well  in the following region of collision energy: 4.4--5 GeV,  10.8--12 GeV and 14.3--17.3 GeV.
In other words, within the present meta-analysis the region of energy $\sqrt{s_{NN}} =$  4.2--4.87 GeV which we associated  above with the hadron-QGP mixed phase  remains almost the same, giving us the overlapping region of energies 
$\langle \sqrt{s_{NN}}\rangle_{overlap} =$ 4.4--4.87  GeV.
However, instead of a single region of reentering the mixed phase now we found two ones.  The most peculiar thing is that at the collision energies $\sqrt{s_{NN}}\,=$  11.3--17.3 GeV the  weighted averaging of HG models demonstrates  somewhat better description of the data than the one of QGP models (see Fig.~\ref{fig6A}). 

How can one understand these results? Our interpretation is as follows. We used two entirely different ways of  averaging in order to study the stability of the results, and hence the most probable results of KTBO-plots 1and 2  are the ones which  coincide.  Therefore, the most probable energy range for the second  mixed phase region  is  $\langle \sqrt{s_{NN}}\rangle_{overlap} =$ 10.8--12 = $11.4 \pm 0.6$  GeV. 
\begin{table}[t]
\begin{tabular}{|c|c|c|c|c|} \hline                
 &  \multicolumn{2}{c|}{Arithmetic averaging}   &   \multicolumn{2}{c|}{Weighted averaging} \\ \hline
$\sqrt{s_{NN}}$ (GeV)      & HG models & QGP models & HG models & QGP models   \\ \hline
%%%
     3.1           & $1.85 \pm 0.575$ & $4.1 \pm 1.84$  & $0.104 \pm 0.077$ &  $ 1.62\pm0.49 $ \\   \hline
     3.6           & $0.913 \pm 0.442$ & $ 3.976\pm1.725 $ & $0.272 \pm 0.1$ &  $ 0.94\pm0.395 $ \\   \hline
     4.2           & $2.086 \pm 0.762$ & $2.616\pm 1.306$ & $0.1042 \pm 0.08$ &  $ 1.14\pm0.46 $ \\   \hline     
     4.9           & $2.813 \pm 0.9$ & $3.57 \pm 0.608$ & $1.086 \pm 0.13$ &  $0.772 \pm 0.22$ \\   \hline
     5.4          & $6.1 \pm 0.854$ &  N/A & $1.733 \pm 0.346$ &  N/A  \\   \hline
     6.3           & $7.24 \pm 2.06$ & $1.78 \pm 0.37$ & $6.044 \pm 1.737$ &  $0.493 \pm 0.54$ \\   \hline
     7.6           & $5 \pm 0.9$ & $1.08 \pm 0.27$ & $2.4 \pm 0.358$ &  $0.405 \pm 0.126$ \\   \hline
     8.8           & $3.66 \pm 1.06$ & $1.41 \pm 0.613$ & $2.355 \pm 0.208$ &  $0.53 \pm 0.1$ \\   \hline
     12.3           & $2.29 \pm 0.9$ & $2.38 \pm 0.33$ & $0.0564 \pm 0.027$ &  $0.6 \pm 0.11$ \\   \hline
     17.3          & $10 \pm 1.51$ & $3.475 \pm 0.427$ &  $0.38 \pm 0.0531$ &  $0.434 \pm 0.12$ \\   \hline                                                                               
                \end{tabular}
\caption{The resulting QDD values  provided by two ways of averaging.  These results of HG and QGP  models are shown in KTBO-plots 1 and 2.}
\label{table2}
\end{table}

As an independent check up, let us consider the energy ranges $\sqrt{s_{NN}} = \{11.75 \pm 1.75;  11.4 \pm 0.6; 15.5\pm 1.5\}$ GeV of the second mixed phase region which we found from the KTBO-plots 1 and 2  as the results of independent meta-measurements.  Then applying to them the both ways of averaging, i.e.  using Eqs. (\ref{EqIV}) and (\ref{EqVI}), we obtain  
\begin{eqnarray}\label{EqXXI}
\langle \sqrt{s_{NN}} \rangle_{aa} = 12.9 \pm 0.8~{\rm  GeV}\,, \quad \langle \sqrt{s_{NN}} \rangle_{wa} = 11.95 \pm 0.53~{\rm  GeV}\,.
\end{eqnarray}
The weighted  averaging in (\ref{EqXXI}) gives us nearly the same estimate which  we found from the overlapping regions 
in  the KTBO-plots 1 and 2, while the arithmetic averaging provides us with a more conservative estimate. 
We suggest to use this value $\langle \sqrt{s_{NN}} \rangle_{aa} = 12.9 \pm 0.8$ GeV as the most probable estimate 
for the (tri)critical endpoint collision energy.  The  main reason is that the two other estimates, i.e.  $\langle \sqrt{s_{NN}}\rangle_{overlap} = 11.4 \pm 0.6$  GeV and  $\langle \sqrt{s_{NN}} \rangle_{wa} = 11.95 \pm 0.53$ GeV,  correspond
to the second mixed phase region and, hence, it is logical to assume  that the (tri)critical endpoint energy is located more close to the upper 
boundary of these estimates or even at  a slightly higher energy. 
This is exactly the region provided by the arithmetic averaging of
the collision energy values.  Also such a conclusion is supported by the fact that the endpoint was, so far, not found at 
$\sqrt{s_{NN}} =$  11.5 GeV and $\sqrt{s_{NN}} =$  12.3 GeV.  
Recall also   the main conclusion of RHIC Beam Energy Scan  \cite{Mohanty13} that the properties
of strongly interacting  matter created  in  A+A collisions at energies below and above 
$\sqrt{s_{NN}}\,=\,11.5$ GeV are different. 
Hence, the value  $\langle \sqrt{s_{NN}} \rangle_{aa} = 12.9 \pm 0.8$ GeV is the  present days best estimate which is  provided by the suggested meta-analysis. Hopefully,  it can be improved further, if one accounts for  the RHIC Beam Energy Scan results measured at  the  collision energies  
$\sqrt{s_{NN}} = 11.5$ GeV and  $\sqrt{s_{NN}} = 19.6$ GeV.

\section{Conclusions and Perspectives}

Here we performed the meta-analysis of  the QDD of  the existing A+A event generators without and with QGP existence  which allow one to extract physical information of principal importance. 
These kinds of generators are, respectively, called the HG and QGP model.  A priori we assumed that, despite their imperfectness,  these models contain the grain of truth on the  QGP 
formation in central nuclear collisions which we would like to distillate using the suggested meta-analysis. 
For each collision energy we consider the set of  QDD $\langle\chi^2/n\rangle^h_A\biggl|_{M}$
of  the experimental quantity $A$ of  hadron $h$ described  by the model $M \in \{HG; QGP \}$ as the results of independent meta-measurements. The studied experimental quantities include the transverse  mass spectra, the longitudinal rapidity distributions and yields measured at midrapidity and/or in the full solid angle. 
{In this work we analyzed the strange hadrons only, since they provide us with one of  the most popular probes  both of   
experimental measurements and of theoretical investigations.}
Using two ways of averaging for the 
QDD $\langle\chi^2/n\rangle^h_A\biggl|_{M}$ and its error, we were able to extract the QDD by two kinds of models at each collision energy.  

Comparing the results found by these two kinds of models we were able to locate the regions of their equal QDD  at the collision energies $\langle \sqrt{s_{NN}}\rangle_{overlap} =$ 4.4--4.87 GeV and  $\langle \sqrt{s_{NN}}\rangle_{overlap} =$ 10.8--12 GeV, which we identified with the mixed phase regions. As expected, at center of mass energies below 4.2 GeV the HG models `work' better than the QGP ones. As it is seen from the KTBO-plots 1 and 2, in the collision energy range $\sqrt{s_{NN}}\,=$ 5--10.8 GeV the HG models  fail to reproduce the vast majority of  data, while the QGP models reproduce  data rather well. Therefore, this region we associated with  QGP. 
Unfortunately,   results of the  A+A event  generators used  here did not allow us yet to uniquely  interpret 
our  findings at the collision energies  above 12 GeV.  Our educated guess is that the energy range $\langle \sqrt{s_{NN}} \rangle_{aa} = 12.9 \pm 0.8$ GeV corresponds to the vicinity of the (tri)critical endpoint. Such a guess is, on the one hand,  supported by the phenomenological arguments of  Ref.  \cite{Quarkyonic} on  the location of (tri)critical endpoint. On the other hand, our hypothesis is also  confirmed 
by the results of lattice QCD  \cite{ChirSusc1,ChirSusc2, ChirLim1,ChirLim2} that the chemical freeze-out states at 
$\sqrt{s_{NN}} = $ 17.3 GeV belong to the cross-over region and, hence, the (tri)critical endpoint should be located at lower energies of collision.  

Of course,  the energies $\langle \sqrt{s_{NN}} \rangle_{aa} = 12.97 \pm 0.92$ GeV may  correspond to the second 
phase transition of  QCD,  but even in this case the (tri)critical endpoint of the QCD phase diagram should be located nearby.
{Perhaps,  our findings may help to interpret the conclusions of the RHIC Beam Energy Scan Program \cite{Mohanty13} that below and above $\sqrt{s_{NN}}\,=\,11.5$ GeV one, respectively,  probes the different phases of  QGP  which have different properties.
We hope that  this question could  be answered, if the RHIC Beam Energy Scan data measured at  center of mass energies 11.5 GeV and 19.7 GeV were analyzed by the event generators and then  reanalyzed using  the present approach. 
We would like to stress that the present work strongly supports conclusions of the generalized shock adiabat model 
\cite{Bugaev:2014a,Bugaev:2014b} that the hadron-QGP mixed phase is located between the center of mass energies 
4.2 GeV and 4.87 GeV.  On the other hand, there is absolutely no evidence for the onset of deconfinement at 
the center of mass collision energy 7.6 GeV, as it is claimed in  \cite{Horn,Step,Gazd_rev:10}. 
Therefore, we believe that 
without establishing a much more convincing  relation  between the irregularities discussed in \cite{Gazd_rev:10} and the deconfinement phase transition one simply cannot consider  them to be signals of deconfining transition. 
 }

{The current set of analyzed models does not cover the full spectrum of  existing ones, but already  this sample 
was sufficient  to obtain interesting results and show the power of the suggested meta-analysis.}
Also it is worth to note that to essentially  change  the results of present  meta-analysis  it would be necessary to greatly improve   the description of  existing data  either by HG models in the center of mass collision energy range 5.4--12.3 GeV or by 
QGP models at energies below 4.2 GeV. Otherwise, it would be hard to change the obtained results, since they are based on the analysis of  more than  100 different experimental data sets.

Due to  absence of necessary information on the QDD  of A+A collisions, we were forced to make the  tantalizing efforts in order to scan  both the experimental data and their description by the event generators of such collisions from the published figures. Such efforts could be avoided, if all existing experimental data  on A+A collisions were collected together 
like it is done in the Particle Data Group  for collisions of elementary particles.  Also the  success of  present meta-analysis suggests that  the results  of theoretical works which  describe  data  should be presented not only in figures, but also they should be available  as the numbers in the form of  QDD $\langle\chi^2/n\rangle^h_A\biggl|_{M}$  and as the total mean deviation squared  per degree of freedom (if such a number can be found).  Then the first of them can be used for the meta-analysis we performed here, while the second ones will help us to determine the most successful A+A event generators
of available data.  Such an information can be used for planning the collision energy  of  ongoing experiments and 
for further  improvement of the existing A+A event generators.  We believe that only along this way the heavy ion collisions community will be able to solve its ultimate tasks, i.e. to locate the hadron-QGP mixed phase and to  discover the QCD phase diagram endpoint.  Therefore, as the  first step in this direction in Tables I and  III-XI we present the  values of  QDD   used in our meta-analysis for the arithmetic averaging  along with the final values for the both ways of averaging  used in the KTBO-plots 1 and 2 (see Table II).  The values of  QDD 
obtained  for the weighted averaging will be published elsewhere. 

As the first practical steps we suggest to the RHIC Beam Energy Scan Program  to make a few measurements in the center of mass collision energy ranges from 3.9 GeV to 5.15 GeV and from 10.8 GeV to 15.5 GeV.  The first of them may lead to a discovery of the hadron-QGP mixed phase, while the second one may lead to a location of the QCD phase diagram endpoint.\\

{\bf Acknowledgements.} The authors would like to thank M. Bleicher, A.I. Ivanytskyi,   L.M. Satarov and A. Taranenko for fruitful discussions. 
Special thanks go to D. H. Rischke for  his  valuable comments. 
This work was supported in part by the Program of Fundamental Research of the Department of Physics and Astronomy of NAS of  Ukraine. K.A.B. acknowledges a partial support provided by the Helmholtz International Center for FAIR within the framework of the LOEWE program launched by the State of Hesse.

\vspace*{1.1mm}

\section{Appendix: Description of the models compared}

{
Here we  briefly discuss  characteristics of the models used in the present meta-analysis. 
We have to apologize  for giving a few basic references only.  The main criterion for choosing these models
was   the wide acceptance of their results by heavy ion community and a requirement  that their results are 
well documented.  In our analysis we  on purpose included not only cascades, hadronic or partonic, but also  statistical  (SHM)
and hydrodynamical  (3FD) models, since we believe that  the meta-analysis    works better, if  the models of different kinds are
included.  
We  consider a  model to be of  a  QGP  type,
if  in it, explicitly or implicitly, there appears   a dense medium of quarks and gluons (partons), whose properties are entirely different from the ones of  hadronic medium employed  in this  model.

%\subsection{ARC}
{\bf ARC} (A Relativistic Cascade) \cite{Arc,Schlagel:1992fh} is a purely hadronic cascade, developed in the Brookhaven National Laboratory  in the early 90s.  ARC is designed for and limited to AGS energies. It does not include any potentials, mean field effects, off-shellness, strings or in-medium effects. Hadrons participate in $2 \to 2$ elastic and inelastic collisions, resonance formations and decays. Cross-sections and their parameterizations are taken from CERN-HERA and LBL compilations as well as from papers 
%%%by Blobel \emph{et al} and Rossi \emph{et al}
 \cite{ARC_cross_sections}. Collisions  are triggered using a  geometrical criterion: if distance between particles $r_{ij} \le (\sigma_{ij}/\pi)^{1/2}$, where $\sigma_{ij}$ is corresponding total cross-section, then collision happens. Formation time of 1 fm/c is introduced. Apparently, this is a HG model.

%\subsection{RQMD}

{\bf RQMD} (Relativistic Quantum Molecular Dynamics) \cite{Sorge:1989vt,Sorge:1995dp} is a hadron cascade, developed since 1989 in the Institute of Theoretical Physics of the University of  Frankfurt upon Main. It employs  nuclear potentials, implemented in a Lorentz-covariant manner, at low energies and fills cross-section with interacting strings at higher energies. Interactions of energy up to 2-3 GeV mainly proceed through hadronic resonances, such as $\rho$, $\eta$, $K^*$, $\phi$, $\Delta$, $N'$, parametrized by many-channels Breit-Wigner formula. Unknown cross-sections are calculated using additive quark model, namely, $\sigma_{tot} \sim (1-0.4 x_1)(1- 0.4x_2)$, where $x_{1,2} = \frac{N_s}{N_u+N_d}$ - strange to non-strange quarks ratio in the corresponding hadron. Like in ARC, there is a formation time of 1 fm/c and geometrical collision criterion. Hadronic cross-sections are taken from different sources, compared to ARC \cite{RQMD_cross_sections}. At higher energies interactions go via strings using phenomenological Lund model, extended by possibility for strings to interact, forming  so-called ropes (see \cite{Bierlich:2014xba} for review). RQMD is often believed  to be valid at a broad collision energies range from $\sim$50 MeV to hundreds GeV per nucleon.

%\subsection{UrQMD}

{\bf UrQMD} (Ultra-relativistic Quantum Molecular Dynamics) is a successor of RQMD, developed  in Frankfurt ITP  in the late 90s \cite{Bass98}. Most recent version can be run as a hybrid model, but we consider only earlier pure hadronic cascade versions. Unlike RQMD, UrQMD employs non-relativistic nuclear potentials, that are switched off  {at relative momentum of colliding hadrons  above 2  GeV.} Resonance table is significantly improved  compared to RQMD. Cross-sections are mainly taken from Particle Data Group (PDG) and CERN-HERA compilations and are  parametrized using multi-channel Breit-Wigner ansatz. Experimentally unknown channels, such as resonance-resonance or hyperon-resonance scattering, are parametrized using phenomenological additive quark model.  As RQMD and HSD, UrQMD employs Lund string model at  center of mass collision energies $\sqrt{s_{NN}} >3$ GeV. Unlike RQMD, UrQMD strings do not interact.

%\subsection{HSD}

{\bf HSD} (Hadron String Dynamics) is a hadron cascade developed at Giessen in the mid-90s \cite{HSDgen,
S5.4a}. It employs potentials, implemented as a relativistic mean-field. HSD particles propagate off-shell and their self-energies are modified in the medium. In the hadronic scattering part HSD is the successor of BUU code,  developed by the Giessen group \cite{Wolf:1993pja}. Collisions happen according to the geometrical criterion, as in ARC and RQMD. Cross-sections are taken from similar sources to ARC plus CERN ISR compilation. In some cases strange particle cross-sections are re-parametrized to change cross-section behavior in the energy regions, where data are  not available \cite{Cassing:1996xx}. At {the  center of  mass} collision energies of nucleons  higher  than 2.6 GeV interaction  proceeds  via strings \cite{HSDgen}, operating according to Lund string model. Strings cannot interact: ropes and string fusion like in RQMD are not implemented. Parameters of Lund model are changed from energy to energy to match experimental data \cite{S5.4a}.

%\subsection{AGSHIJET}

{\bf AGSHIJET} \cite{HIJETa} is a  hadron cascade inheriting from HIJET \cite{Shor:1988vk}, which, in turn, comes from a ISAJET generator for $p+p$ collisions  \cite{Paige:1986kt}.  All these models were developed at Brookhaven National Laboratory  in late 80s - early 90s.  AGSHIJET  operates with strings and uses not the Lund model, but a Field-Feynmann model, which has a different parametrization for fragmentation functions. HIJET simulates A+A collision as a sequence of independent $NN$ collisions, where individual $NN$ collisions are treated using ISAJET. Further HIJET applies a  hadron cascade, similar to ARC, to the obtained products. In the AGSHIJET resonance treatment was improved, $N^*$ added and Field-Feynmann model parameters were adjusted to reproduce $p+p$ reaction  cross-section at laboratory  frame momentum below 50 GeV. 

%\subsection{SHM}

{\bf SHM} (Statistical Hadronization Model) is not a microscopic, but a pure statistical model of hadron gas \cite{Becattini:2003wp,SHMgen}, which allows one  to obtain hadron yields (but not spectra) under the assumption of a sharp chemical freeze-out. In this model 
it is assumed that before chemical freeze-out matter is in thermal and chemical equilibrium and after it inelastic reactions are frozen  and only resonance decays and elastic reactions occur. Thermal yields of hadrons  are calculated from the equilibrium ideal gas partition function and read as $N = \frac{g V |\gamma_s|^{n_s}}{2\pi^2} \int \frac{k^2 dk}{e^{(E-\mu)/T} \pm 1}$. Temperature $T$, volume $V$, baryon chemical potential $\mu$ and strangeness suppression factor $\gamma_s$ are free parameters, adjusted at each collision energy to provide the best fit of experimental hadronic  yields. Interaction between hadrons is taken into account via adding resonances as independent components of ideal gas.

%\subsection{QuarkComb}

{\bf QuarkComb} (Quark Combination model) \cite{QComb} is a  phenomenological gluonless model, which generates effective  quarks with rapidity distribution of $\frac{dN_q}{N_q \, dy} = B \left( e^{-|y|^a/2\sigma^2} - e^{-|y_{beam}|^a/2\sigma^2}\right)$, where $B$ is normalization constant, while $a$ and $\sigma$ are adjustable parameters. Number and flavour of quarks are generated according to a phenomenological model of hadronization introduced by Xie and Liu in 1988 \cite{Xie:1988wi} and tuned to describe hadron production in $e^++ e^-$ collisions. Effective  (anti)quarks are sorted in rapidity and neighbors in the array are combined into mesons or baryons: $qq\bar{q}$ and its permutations give a meson and one quark remains in the array, while $qqq$ gives a baryon. To obtain final spectra hadron resonances are decayed. According to our criterion  this is QGP model.

%\subsection{3FD}

{\bf 3FD} (3-fluid dynamics) is a hydrodynamical model of heavy ion collision, developed by Ivanov and Russkikh \cite{3FDgen}. It considers ion collision as three fluids: two baryon-rich fluids corresponding to projectile and target and the third one, corresponding to produced particles. A following system of  coupled equations   of  perfect  fluid dynamics 
\begin{eqnarray}\label{EqAI}
  \partial_{\mu} T^{\mu\nu}_p &=& -F_p^{\nu} + F_{fp}^{\nu} \,, \\
  \partial_{\mu} T^{\mu\nu}_t &=& -F_t^{\nu} + F_{ft}^{\nu} \,,\\
  \partial_{\mu} T^{\mu\nu}_f &=& F_p^{\nu} - F_{fp}^{\nu} + F_t^{\nu} - F_{ft}^{\nu} \,,
\end{eqnarray}
is solved for the equation of state with the first order phase transition between hadronic matter and QGP  \cite{3FDII,3FDIII}.
The left hand side  of system  (\ref{EqAI})  is standard for relativistic hydrodynamics, while its  right  hand side  represents interaction between fluids. Since $\partial_{\mu} (T^{\mu\nu}_p + T^{\mu\nu}_t + T^{\mu\nu}_f) = 0$, total energy and momentum are conserved.  Freeze-out is performed using a modified Cooper-Frye procedure  and, hence,  it experiences 
the typical problems of other hydrodynamic models   discussed  in  \cite{Freeze:96}. 
After freeze-out the obtained resonances are decayed.  
For considered two phase  equation of state we assign  3FD approach  to  QGP models. 

%\subsection{PHSD}

{\bf PHSD} (Parton-Hadron-String Dynamics) \cite{PHSDgen} is the most advanced  cascade model based on  HSD.
It combines  the HSD hadronic sector with a new model for partonic transport and hadronization, so-called DQPM (Dynamical Quasi-Particle Model). DQPM describes QCD properties in terms of single-particle Green'€™s functions (in the sense of a two-particle irreducible approximation). Elastic and inelastic $q+q \leftrightarrow q+q$, $g+g \leftrightarrow g+g$, $g+g \leftrightarrow g$, $q+\bar{q} \leftrightarrow g$ reaction are included, fulfilling the detailed balance condition. Hadronization is occurring continuously based on  Lorentz-invariant transition rates. Obtained hadron is embedded to the HSD cascade. If the hadron mass is too large, then it is considered as a HSD string. Apparently, this is QGP model.

%\subsection{Core-Corona Model}

{\bf Core-Corona Model} \cite{CoreCorMa} is a phenomenological approach  suggested  in 2005 to parameterize and to represent  experimental data in a simple, but effective way. It  was further  developed in  \cite{CoreCorMb}. This model is based  on the assumption that the fireball created in nuclear collisions is composed of a core, which has the same properties as a very central collision system, and a corona, which has different properties and is considered as  a superposition of independent nucleon-nucleon interactions. The sizes of core and corona are  defined by the condition that the corona is formed by those nucleon-nucleon collisions in which both participants  interact only once during the entire  process of A+A collision. The fraction of single scatterings is calculated using straight line geometry as described in the Glauber model \cite{Glauber}.
Since nowadays  there are no doubts that at sufficiently high energies in central A+A collisions the QGP is formed, 
therefore at such  energies 
we regard  a core state of  this  model as an implicit parameterization  of QGP state.  
}

%\begin{multicols}{2}
%--------------------   Bibliography  ------------------------------------%
\vspace{3mm}
\begin{center}

\small{}
\end{center}

%%%\newpage

%%%\input{AAtables_new.tex}

\begin{table}[t]
\begin{tabular}{|c|c|c|c|} \hline                    
                      &  \multicolumn{3}{c|}{$\sqrt{s_{NN}}\,=$  3.1 GeV}      \\ \hline
                   & $m_{T}$-distribution & rapidity distribution & Yields       \\ \hline
$\langle\chi^2/n\rangle =$     &                &                   & $0.856\pm0.755$ $\left(\frac{d N}{d y}\bigl|_{y=0}\&\quad4\pi \right)$ \\
$\Lambda(\Sigma^{0})$ &   N/A       &       N/A         & HSD \& UrQMD1.3(2.1)  \\
                   &                &                   & Figs. 1,  Ref. \cite{S4.9KL}     \\ \hline
$\langle\chi^2/n\rangle =$     & $1.575\pm0.724$  &                   & $5.9\pm4.85$    \\
$K^{+}$ set1       & 3FD            &        N/A        &  3FD   \\
                   & Fig. 1,  Ref. \cite{3FDIII} &        &   Fig. 9,  Ref. \cite{3FDII}   \\  \hline 
$\langle\chi^2/n\rangle =$     & $1.49\pm0.676$  &                   & $7.43\pm5.45$    \\
$K^{-}$ set1       & 3FD            &        N/A        &  3FD   \\
                   & Fig. 1,  Ref. \cite{3FDIII} &        &   Fig. 9,  Ref. \cite{3FDII}   \\  \hline
$\langle\chi^2/n\rangle =$     & $1.35\pm0.476$ &                   & $8.246\pm3.316$ $\frac{dN}{dy}\bigl|_{y=0}$ \\
$K^{+}$ set 2      & HSD \& UrQMD2.0&       N/A         & HSD \& UrQMD1.3(2.1)  \\
                   &                &                   & Fig. 1 or 2,  Ref. \cite{S4.9KL}   \\ \hline
$\langle\chi^2/n\rangle =$     & $1.7\pm0.51$   &                   & $0.0714\pm0.31$ $\frac{dN}{dy}\bigl|_{y=0}$  \\
$K^{-}$ set 2      & HSD \& UrQMD2.0 &     N/A         & HSD \& UrQMD1.3(2.1)      \\
                   &                &                   & Figs.1 or 2,  Ref. \cite{S4.9KL}  \\   \hline
$\langle\chi^2/n\rangle =$     & $1.525\pm0.35$ &                   &               \\
$K^\pm$            & HSD \& UrQMD2.0 &       N/A         &     N/A        \\
                   & Fig. 7,  Ref. \cite{S4.9KL}&         &               \\ \hline                               

\end{tabular}
\caption{Same as in Table \ref{tableI}, but for collision energy $\sqrt{s_{NN}}\,=$  3.1 GeV.
}
\label{tableII}
\end{table}
%
%}
%{
\begin{table}[t]
\begin{tabular}{|c|c|c|c|} \hline 
                                   
                      &  \multicolumn{3}{c|}{$\sqrt{s_{NN}}\,=$  3.6 GeV}      \\ \hline
                   & $m_{T}$-distribution & rapidity distribution & Yields       \\ \hline
$\langle\chi^2/n\rangle =$     &                &                   & $0.862\pm0.76$ $\left(\frac{d N}{d y}\bigl|_{y=0}\&\quad4\pi \right)$ \\
$\Lambda(\Sigma^{0})$ &     N/A     &       N/A         & HSD \& UrQMD1.3(2.1)  \\
                   &                &                   & Figs. 1,  Ref. \cite{S4.9KL}     \\ \hline
$\langle\chi^2/n\rangle =$     & $5.53\pm2.74$  &       & $2.13\pm1.7$ $\left(\frac{dN}{dy}\bigl|_{y=0} \right)$ \\ 
$K^{+}$ set 1      & $m_{T}$-distr.+Yield:3FD &       N/A         & HSD \& UrQMD1.3(2.1)  \\   
                   & $m_{T}$:Fig.1,Ref.\cite{3FDIII};Yield:Fig.9,Ref.\cite{3FDII} &     &   Fig. 1 or 2,  Ref. \cite{S4.9KL} \\ \hline
$\langle\chi^2/n\rangle =$     & $1.17\pm0.46$  &                   &          \\
$K^{+}$ set 2       & HSD \& UrQMD2.0&       N/A          &   N/A     \\
                   & Fig. 7,  Ref. \cite{S4.9KL}&         &       \\ \hline 
$\langle\chi^2/n\rangle =$     & $0.453\pm0.286$ &                & $0.0824\pm0.332$ $\left(\frac{dN}{dy}\bigl|_{y=0} \right)$ \\              $K^{-}$  set1  & HSD \& UrQMD2.0 &      N/A            & HSD \& UrQMD1.3(2.1)  \\
                   & Fig. 7,  Ref. \cite{S4.9KL}&         & Fig.  1 or 2,  Ref. \cite{S4.9KL} \\ \hline
$\langle\chi^2/n\rangle =$     & $0.494\pm0.424$  &                   &    $4.35\pm4.17$      \\
$K^{-}$  set2      &   3FD            &       N/A         &  3FD   \\
                   & Fig. 1,  Ref. \cite{3FDIII} &        &   Fig. 9,  Ref. \cite{3FDII}   \\  \hline
                      &  \multicolumn{3}{c|}{$\sqrt{s_{NN}}\,=$  4.2 GeV}      \\ \hline
                   & $m_{T}$-distribution & rapidity distribution & Yields       \\ \hline
$\langle\chi^2/n\rangle =$     &                &                   & $3.35\pm1.47$ $\left(\frac{d N}{d y}\bigl|_{y=0}\&\quad4\pi \right)$ \\
$\Lambda(\Sigma^0)$ &     N/A       &      N/A          & HSD \& UrQMD1.3(2.1)    \\
                   &                &                   & Fig.  1,  Ref. \cite{S4.9KL}     \\ \hline
$\langle\chi^2/n\rangle =$     & $0.81\pm0.3$   &                   & $0.8343\pm0.75$ $\left(\frac{dN}{dy}\bigl|_{y=0} \right)$ \\
$K^\pm$            & HSD \& UrQMD2.0 &      N/A         & HSD \& UrQMD1.3(2.1)  \\
                   & Fig. 7,  Ref. \cite{S4.9KL}&        & Fig.  1 or 2,  Ref. \cite{S4.9KL} \\ \hline
$\langle\chi^2/n\rangle =$     & $3.367\pm1.22$   &        &  $5.7\pm4.78$   \\ 
$K^{+}$             & 3FD            &        N/A       &  3FD   \\
                   & Fig. 1,  Ref. \cite{3FDIII} &        &   Fig. 9,  Ref. \cite{3FDII}   \\  \hline
$\langle\chi^2/n\rangle =$     & $0.73\pm0.54$   &        &  $.665\pm1.63$   \\
$K^{-}$                   & 3FD            &        N/A       &  3FD   \\
                    & Fig. 1,  Ref. \cite{3FDIII} &        &   Fig. 9,  Ref. \cite{3FDII}   \\  \hline
\end{tabular}
\caption{Same as in Table \ref{tableI}, but for collision energies $\sqrt{s_{NN}}\,=$ 3.6 and 4.2 GeV.
}
\label{tableIII}
\end{table}
%}

%{
%
\begin{table}[t]
\begin{tabular}{|c|c|c|c|} \hline 
                       &  \multicolumn{3}{c|}{$\sqrt{s_{NN}}\,=$  5.4 GeV}      \\ \hline
                   & $m_{T}$-distribution & rapidity distribution & Yields       \\ \hline
$\langle\chi^2/n\rangle =$    &         & $8.17\pm2$              &  \\ 
$K^\pm$           &    N/A  & HSD                     &  N/A    \\ 
                  &         & Fig. 14,  Ref. \cite{S5.4a}  &   \\ \hline 
$\langle\chi^2/n\rangle =$    &         & $0.45\pm0.6,$ $1.64\pm1.14$  & $1.1\pm2.1,$ $1.35\pm2.3$ \\ 
 $\Lambda$ set 1  &    N/A     &ARC(SiSi(Pb))                 &  ARC(SiSi(Pb))   \\ 
                   &         &Figs. 4-5,  Ref. \cite{S5.4b} &    Ref. \cite{S5.4b}       \\ \hline 
$\langle\chi^2/n\rangle =$    &         & $2.24\pm1.3,$ $6.1\pm2.2$  & $8.9\pm6,$ $10.2\pm6.4$ \\ 
 $\Lambda$ set 2  &      N/A   &AGSHIJET-N*(SiSi(Pb))      &  AGSHIJET-N*(SiSi(Pb))   \\ 
                   &         &Figs. 4-5,  Ref. \cite{S5.4b} &    Ref. \cite{S5.4b}       \\ \hline
$\langle\chi^2/n\rangle =$    &         & $11.7\pm3,$ $22\pm4$  & $1\pm2,$ $0.3\pm1$ \\ 
 $K_{S}^{0}$         &    N/A     &ARC(SiSi(Pb))                 &  ARC(SiSi(Pb))   \\ 
                   &         &Figs. 2-3,  Ref. \cite{S5.4b} &    Ref. \cite{S5.4b}       \\ \hline 
$\langle\chi^2/n\rangle =$    &         & $1.06\pm0.9,$ $2.4\pm1.4$  & $3.5\pm3.7,$ $7\pm5$ \\ 
 $\Lambda$ set 3  &    N/A     &AGSHIJET-N*(SiSi(Pb))      &  AGSHIJET-N*(SiSi(Pb))   \\ 
                   &         &Figs. 2-3,  Ref. \cite{S5.4b} &    Ref. \cite{S5.4b}       \\ \hline        

\end{tabular}
\caption{Same as in Table \ref{tableI}, but for the  collision energy $\sqrt{s_{NN}}\,=$ 5.4 GeV. The notation 
`ARC(SiSi(Pb))'  used for $\Lambda$ hyperons (set 1) of $y$-distribution means  that  at preceding  row of this column we give  two values of $\langle\chi^2/n\rangle$ which correspond to Si+Si and Si+Pb collisions calculated for the  event generator ARC. 
}
\label{tableV}                    
\end{table}
%}
%{
%
\begin{table}[t]
\begin{tabular}{|c|c|c|c|} \hline
                      &   \multicolumn{3}{c|}{$\sqrt{s_{NN}}\,=$ 6.3 GeV}      \\  \hline
                    & $m_{T}$-distribution & rapidity distribution & Yields       \\ \hline
$\langle\chi^2/n\rangle =$     & $1.215\pm0.73$ & $2.28\pm0.522$  &             \\
$K^\pm$            & $m_{T}$-distr.+Yield:3FD   & QuarkComb. model &    N/A     \\
                   & $m_{T}$:Fig.1,Ref.\cite{3FDIII};Yield:Fig.9,Ref.\cite{3FDII} & Fig. 2  Ref. \cite{QComb}&          \\  \hline
$\langle\chi^2/n\rangle =$     &               & $1\pm0.72$        & $2.91\pm3.41,$ $8.2\pm5.73$     \\
$\phi$             &     N/A       & QuarkComb. model & SHM, UrQMD     \\
                   &               & Fig. 2  Ref. \cite{QComb}&  Fig. 17  Ref. \cite{Blume}           \\ \hline
$\langle\chi^2/n\rangle =$     &               & $5.645\pm1.5$     &       \\
$\Lambda$          &     N/A       & QuarkComb. model &     N/A  \\
                   &               & Fig. 2,  Ref. \cite{QComb} &  \\  \hline
$\langle\chi^2/n\rangle =$     &               & $0.266\pm0.516$   &       \\
$\bar{\Lambda}$    &     N/A       & QuarkComb. model &   N/A    \\
                   &               & Fig. 2,  Ref. \cite{QComb} & \\   \hline
$\langle\chi^2/n\rangle =$     &               & $0.277\pm0.39$    &    \\
$\Xi^{-}$          &    N/A        & QuarkComb. model &   N/A \\
                   &               & Fig. 2,  Ref. \cite{QComb} &  \\  \hline

\end{tabular}
\caption{Same as in Table \ref{tableI}, but for the  collision energy $\sqrt{s_{NN}}\,=$ 6.3 GeV.
}
\label{tableVI}
\end{table}
%}
%{
%
\begin{table}[t]
\begin{tabular}{|c|c|c|c|} \hline
                      &   \multicolumn{3}{c|}{$\sqrt{s_{NN}}\,=$ 7.6 GeV}      \\  \hline 
                   & $m_{T}$-distribution & rapidity distribution & Yields       \\ \hline
$\langle\chi^2/n\rangle =$     & $1.51\pm0.53$  & $0.6\pm0.265$     &                      \\
$K^\pm$            & 3FD           & QuarkComb. model &        N/A             \\
                   & Fig. 1, Ref. \cite{3FDIII}& Fig. 2  Ref. \cite{QComb} & \\  \hline
$\langle\chi^2/n\rangle =$     &               & $1.413\pm0.752$   & $0.567\pm1.5,$ $1.44\pm2.4$ \\
$\phi$             &      N/A      & QuarkComb. model & SHM,UrQMD        \\
                   &               & Fig. 2,  Ref. \cite{QComb} & Fig. 17,  Ref. \cite{Blume}   \\  \hline
$\langle\chi^2/n\rangle =$     &               & $0.4\pm0.4$       &       \\
$\Lambda$          &       N/A     & QuarkComb. model &   N/A   \\
                   &               & Fig. 2,  Ref. \cite{QComb}   &      \\  \hline
$\langle\chi^2/n\rangle =$     &               & $0.363\pm0.5$     &  \\
$\bar{\Lambda}$    &     N/A     & QuarkComb. model &  N/A \\
                   &               & Fig. 2,  Ref. \cite{QComb}  &   \\  \hline
$\langle\chi^2/n\rangle =$     &               & $2.16\pm0.856$    &  \\
$\Xi^\pm$          &     N/A       & QuarkComb. model &   N/A \\
                   &               & Fig. 2,  Ref. \cite{QComb} &    \\   \hline

\end{tabular}
\caption{Same as in Table \ref{tableI}, but for the  collision energy $\sqrt{s_{NN}}\,=$ 7.6 GeV.
}
\label{tableVII}
\end{table}
%}
%{
%
\begin{table}
\begin{tabular}{|c|c|c|c|}   \hline
                      &    \multicolumn{3}{c|}{$\sqrt{s_{NN}}\,=$ 8.8 GeV}    \\  \hline
                   & $m_{T}$-distribution & rapidity distribution & Yields       \\ \hline
$\langle\chi^2/n\rangle =$     & $4.03\pm0.9$  & $2.76\pm0.56,$ $3.826\pm0.654$ &  $0.1\pm0.3$       \\
$K^\pm$  set 1     & 3FD           & HSD, PHSD                       &   Core Corona        \\
                   & Fig.1,Ref. \cite{3FDIII} & Fig. 15  Ref. \cite{Phsd} & Fig. 10 Ref. \cite{CoreCor} \\  \hline
$\langle\chi^2/n\rangle =$    & $2.33\pm0.7$  &  $0.98\pm0.333$    &  $10.7\pm0.7$      \\
$K^\pm$  set 2     & PHSD          & QuarkComb. model & y+Yield: HSD \& UrQMD2.3     \\
                   & Fig. 16  Ref. \cite{Phsd} & Fig. 2  Ref. \cite{QComb} & Figs. 6-7,10  Ref. \cite{CoreCor} \\  \hline    
$\langle\chi^2/n\rangle =$     &               & $1.8\pm0.846$     & $0.94\pm1.94,$ $8.614\pm5.9$  \\
$\phi$             &     N/A       & QuarkComb. model & SHM,UrQMD    \\
                   &               & Fig. 2,  Ref. \cite{QComb} & Fig. 17,  Ref. \cite{Blume}   \\  \hline   
$\langle\chi^2/n\rangle =$     &               & $1.51\pm0.78,$ $0.185\pm0.27;$ $0.6\pm0.5$ &            \\
$\Lambda(\Sigma^0)$ &      N/A     & HSD, PHSD; QuarkComb. model                 &  N/A   \\
                   &               & Fig. 17 Ref. \cite{Phsd}; Fig. 2 Ref. \cite{QComb} &    \\  \hline
$\langle\chi^2/n\rangle =$     &               & $2.535\pm1.126,$ $0.586\pm0.54$   &   \\
$\bar{\Lambda}$    &      N/A      & PHSD, QuarkComb. model              &  N/A \\
                   &               & Fig. 18 Ref. \cite{Phsd}, Fig. 2 Ref. \cite{QComb} &  \\  \hline
$\langle\chi^2/n\rangle =$     &               & $1.14\pm0.7,$ $0.95\pm0.667$  &    \\
$\Xi^\pm$          &    N/A        & PHSD, QuarkComb. model        &    N/A \\
                   &               & Figs. 19-20 Ref. \cite{Phsd}, Fig. 2 Ref. \cite{QComb} &    \\   \hline

\end{tabular}
\caption{Same as in Table \ref{tableI}, but for the  collision energy $\sqrt{s_{NN}}\,=$ 8.8 GeV. 
 The notation `y+Yield'  used  for $K^\pm$ mesons (set 2)  means  that we show the value of  $\langle\chi^2/n\rangle$ averaged over the  y-distribution and the particles yield, first, and then the  results  obtained for positive and negative kaons are arithmetically averaged again.  Note that the results taken  from  Fig. 10 of Ref.  \cite{CoreCor} include the most central collisions with $N_{wound}> 300$.
}
\label{tableVIII}
\end{table}
%}
%{
%
\begin{table}
\begin{tabular}{|c|c|c|c|}   \hline
                     &   \multicolumn{3}{c|}{$\sqrt{s_{NN}}\,=$ 12.3 GeV}   \\  \hline
                   & $m_{T}$-distribution & rapidity distribution & Yields       \\ \hline
$\langle\chi^2/n\rangle =$     & $1.26\pm0.4,$ $5.05\pm0.75$  & $1.3\pm0.4,$ $2.2\pm0.5;$ $0.63\pm0.27$ & \\
$K^\pm$            & PHSD, 3FD          & HSD, PHSD; QuarkComb. model   &     N/A         \\
& Fig. 16 Ref. \cite{Phsd},  & Fig. 15 Ref. \cite{Phsd}; & \\
&Fig. 1, Ref. \cite{3FDIII} & Fig. 2 Ref. \cite{QComb}& \\ \hline

$\langle\chi^2/n\rangle =$     &               & $6.16\pm1.326$   & $1.6\pm2.53,$ $5.264\pm4.6$  \\
$\phi$             &    N/A        & QuarkComb.model & SHM,UrQMD    \\
                   &               & Fig. 2 Ref. \cite{QComb} & Fig. 17 Ref. \cite{Blume}   \\  \hline
$\langle\chi^2/n\rangle =$     &               & $0.553\pm0.526,$ $0.076\pm0.2;$ $1.7\pm0.923$ &   \\
$\Lambda(\Sigma^0)$&      N/A        & HSD,PHSD;QuarkComb.model &   N/A \\
                   &               & Fig. 17 Ref. \cite{Phsd}; Fig. 2 Ref. \cite{QComb} & \\  \hline
$\langle\chi^2/n\rangle =$     &               & $3.2\pm1.46,$ $0.24\pm0.4$  &    \\ 
$\bar{\Lambda}(\bar{\Sigma^0})$ &   N/A     & PHSD,QuarkComb.model &   N/A \\
                            &       & Fig. 18 Ref. \cite{Phsd},Fig. 2 Ref. \cite{QComb} &  \\  \hline
$\langle\chi^2/n\rangle =$     &               & $0.74\pm0.6,$ $1.453\pm0.7$ & \\
$\Xi^\pm$          &    N/A        & PHSD,QuarkComb.model       &  N/A \\
                   &               & Figs. 19-20 Ref. \cite{Phsd}, Fig. 2 Ref. \cite{QComb}  &  \\  \hline

\end{tabular}
\caption{Same as in Table \ref{tableI}, but for the  collision energy $\sqrt{s_{NN}}\,=$ 12.3 GeV.
}
\label{tableIX}
\end{table}
%}
%{
%
\begin{table}
\begin{tabular}{|c|c|c|c|}      \hline
                       &      \multicolumn{3}{c|}{$\sqrt{s_{NN}}\,=$ 17.3 GeV}   \\     \hline
                  & $m_{T}$-distribution & rapidity distribution & Yields       \\ \hline
$\langle\chi^2/n\rangle =$    &  $1.83\pm0.2$ & $25.03\pm0.81$ & $0.6\pm0.44$ $\left(\frac{dN}{dy}\bigl|_{y=0} \& \quad4 \pi\right)$ \\
$K^\pm$  set 1    &  5 versions of HSD\& & 2 versions of HSD\&  & HSD \& UrQMD1.3(2.1) \\
                              & UrQMD2.0(2.1)  & UrQMD2.0(2.1) & \\
                  & Figs. 7,8,10,12 Ref. \cite{S4.9KL} &Figs. 9,11 Ref. \cite{S4.9KL}& Fig.  1 Ref. \cite{S4.9KL} \\   \hline

$\langle\chi^2/n\rangle =$    & $2.675\pm0.667,$ $0.724\pm0.348$ & $4.1\pm0.665,$ $18.11\pm0.7$ & $1.3\pm0.81,$ $0.0165\pm0.19$  \\
$K^\pm$  set 2    & 3FD, PHSD            & PHSD, HSD \& UrQMD2.3         & HSD\&UrQMD2.3,  \\
&Fig.1, Ref. \cite{3FDIII}; & Fig.  15 Ref. \cite{Phsd},  & CoreCorona \\  
  
& Fig.  16 Ref. \cite{Phsd}&  Figs. 6-7 Ref. \cite{CoreCor} & Fig.  10 Ref. \cite{CoreCor} \\  \hline   

$\langle\chi^2/n\rangle =$     &      & $1.53\pm0.78,$ $1.314\pm0.725$ & $0.66\pm0.66$ $\left(\frac{dN}{dy}\bigl|_{y=0} \& \quad4 \pi\right)$ \\
$\Lambda(\Sigma^0)$ set 1 &    N/A     & HSD,PHSD                      & HSD \& UrQMD1.3(2.1)   \\
                   &               & Fig. 17 Ref. \cite{Phsd}        & Fig. 1 Ref. \cite{S4.9KL} \\   \hline  
$\langle\chi^2/n\rangle =$    &                 &           & $0.512\pm1.43,$ $0.44\pm1.327$  \\
$\Lambda(\Sigma^0)$ set 2 &     N/A     &     N/A      & HSD, PHSD   \\
                    &                &               & Fig. 21 Ref. \cite{Phsd} \\  \hline
$\langle\chi^2/n\rangle =$     &               & $10.87\pm2.33,$ $18.57\pm3.05$ & $1.4\pm2.37,$ $0.0415\pm0.4$  \\
$\bar{\Lambda}(\bar{\Sigma^0})$ &    N/A    & HSD, PHSD                 & HSD, PHSD  \\
                   &               & Fig. 18 Ref. \cite{Phsd} & Fig. 21 Ref. \cite{Phsd} \\ \hline
\end{tabular}
\caption{Same as in Table \ref{tableI}, but for collision energy $\sqrt{s_{NN}}\,=$ 17.3 GeV. 
The notation `5 versions of HSD\&UrQMD2.0(2.1)' shown for $K^\pm$ meson (set 1) 
means we give  the values of $\langle\chi^2/n\rangle $
 averaged  over   5 types of HSD parameterization (see Ref. \cite{S4.9KL}) and over 2 versions of UrQMD (see ibid.).
 Note that the results taken from Fig. 21 of Ref.  \cite{Phsd}  and from  Fig. 10 of Ref.  \cite{CoreCor} include the most central collisions with $N_{wound}> 300$.}                   
\label{tableX}
\end{table}

\begin{table}
\begin{tabular}{|c|c|c|c|}      \hline
                       &      \multicolumn{3}{c|}{$\sqrt{s_{NN}}\,=$ 17.3 GeV}   \\     \hline
                  & $m_{T}$-distribution & rapidity distribution & Yields       \\ \hline                   
$\langle\chi^2/n\rangle =$     &               &                & $44.6\pm9.44$  \\
$\phi$             &      N/A      &     N/A        & SHM \& UrQMD   \\
                   &               &                & Fig. 17 Ref. \cite{Blume} \\  \hline      
$\langle\chi^2/n\rangle =$     &               & $9.19\pm1.75,$ $1.98\pm0.812$ & $0\pm0;$ $5.513\pm3.32,$ $2.02\pm2.01$ \\
$\Xi^\pm$          &    N/A        & HSD,PHSD                    & UrQMD with reduced masses;     \\ 
                    &               &                            &  HSD,PHSD \\
                   &               & Figs. 19-20 Ref. \cite{Phsd} & Fig. 2 Ref. \cite{Blume}; Fig. 22 Ref. \cite{Phsd}\\  \hline
$\langle\chi^2/n\rangle =$     &               &                    & $0.34\pm1.166$ \\
$\Lambda+\bar{\Lambda}$ &    N/A    &        N/A        & UrQMD with reduced masses  \\
                   &               &                    & Fig. 2 Ref. \cite{Blume} \\  \hline   
$\langle\chi^2/n\rangle =$     &               &                 & $3.65\pm3.82$  \\
$\Omega^\pm$       &    N/A       &         N/A     & UrQMD with reduced masses  \\
                   &                &               & Fig. 2 Ref. \cite{Blume}  \\   \hline

\end{tabular}
\caption{Continuation of  Table \ref{tableX}  for collision energy $\sqrt{s_{NN}}\,=$ 17.3 GeV. 
Note that the results taken from Fig. 22 of Ref.  \cite{Phsd}  and  from Fig. 2 of Ref.  \cite{Blume}  include the most central collisions with $N_{wound}> 300$.
}
\label{tableXI}
\end{table}

\end{document}